\documentclass[longauth]{aa}
\usepackage{graphicx}
\usepackage{setspace}
\usepackage{txfonts}
\usepackage{hyperref}
\usepackage{natbib,twoopt}
\usepackage{threeparttable}
\bibpunct{(}{)}{;}{a}{}{,}
\begin{document}
\title{Dynamical models to explain observations with SPHERE in planetary systems with double debris belts 
        \thanks{Based on observations collected at Paranal Observatory, ESO (Chile)
Program ID: 095.C-0298, 096.C-0241, 097.C-0865 and 198.C-0209 }}

\author{C. Lazzoni \inst{\ref{inst1} \and \ref{inst2}}
 \and S. Desidera \inst{\ref{inst1}}
 \and F. Marzari \inst{\ref{inst2} \and \ref{inst1}}
 \and A. Boccaletti \inst{\ref{inst3}} 
 \and M. Langlois \inst{\ref{inst21} \and \ref{inst4}}
 \and D. Mesa \inst{\ref{inst1} \and \ref{cile}}
 \and R. Gratton \inst{\ref{inst1}}
 \and Q. Kral \inst{ \ref{inst20}}
 \and N. Pawellek \inst{\ref{inst19} \and \ref{inst17} \and \ref{inst5}}
 \and J. Olofsson \inst{\ref{inst18} \and \ref{inst5}}
  \and M. Bonnefoy \inst{\ref{inst9}} 
   \and G. Chauvin \inst{\ref{inst9}}  
   \and A. M. Lagrange \inst{\ref{inst9}}
 \and A. Vigan \inst{\ref{inst4}}
 \and E. Sissa \inst{\ref{inst1}}
 \and J. Antichi \inst{\ref{inst30} \and \ref{inst1}}
 \and H. Avenhaus \inst{\ref{inst5} \and \ref{inst6}  \and \ref{inst7}}
 \and A. Baruffolo \inst{\ref{inst1}} 
 \and J. L. Baudino\inst{\ref{inst3} \and \ref{inst8}}
\and A. Bazzon \inst{\ref{inst6}}
 \and J. L. Beuzit \inst{\ref{inst9} }
\and B. Biller \inst{\ref{inst5} \and \ref{inst11}} 
 \and M. Bonavita\inst{\ref{inst1} \and \ref{inst11}}
 \and W. Brandner \inst{\ref{inst5}}
\and P. Bruno \inst{\ref{inst40}}
 \and E. Buenzli \inst{\ref{inst6}}
 \and F. Cantalloube \inst{\ref{inst9}}
\and E. Cascone \inst{\ref{inst50}}
 \and A. Cheetham \inst{\ref{inst12}}  
 \and R. U. Claudi \inst{\ref{inst1}}
\and M. Cudel \inst{\ref{inst9}}  
\and S. Daemgen \inst{\ref{inst6}}
\and V. De Caprio \inst{\ref{inst50}}
 \and P. Delorme \inst{\ref{inst9}} 
 \and D. Fantinel \inst{\ref{inst1}}
 \and G. Farisato \inst{\ref{inst1}}
   \and M. Feldt \inst{\ref{inst5}}
   \and R. Galicher \inst{\ref{inst3}}
   \and C. Ginski \inst{\ref{inst13}}
   \and J. Girard \inst{\ref{inst9}}
  \and E. Giro \inst{\ref{inst1}}
    \and M. Janson \inst{\ref{inst5} \and \ref{inst14}}
\and J. Hagelberg \inst{\ref{inst9} }
\and T. Henning \inst{\ref{inst5}}
\and S. Incorvaia \inst{\ref{inst60}} 
\and M. Kasper \inst{\ref{inst9} \and \ref{inst17}}
\and T. Kopytova \inst{\ref{inst5}}
  \and J. Lannier \inst{\ref{inst9}}
   \and H. LeCoroller \inst{\ref{inst4}}
 \and L. Lessio \inst{\ref{inst1}}
 \and R. Ligi \inst{\ref{inst4}}
  \and A. L. Maire \inst{\ref{inst5}}
  \and F. M\'enard \inst{\ref{inst9}}
 \and M. Meyer \inst{\ref{inst6} \and \ref{inst15}}
 \and J. Milli \inst{\ref{inst007}}
\and D. Mouillet \inst{\ref{inst9}}
   \and S. Peretti \inst{\ref{inst12}} 
       \and C. Perrot \inst{\ref{inst3}} 
     \and D. Rouan \inst{\ref{inst3}}
  \and M. Samland \inst{\ref{inst5}} 
 \and B. Salasnich \inst{\ref{inst1}}
  \and G. Salter \inst{\ref{inst4}}
\and T. Schmidt \inst{\ref{inst3}}
\and S. Scuderi \inst{\ref{inst40}}
\and E. Sezestre \inst{\ref{inst9}}
\and M. Turatto \inst{\ref{inst1}} 
\and S. Udry \inst{\ref{inst12}}
\and F. Wildi \inst{\ref{inst12}}
    \and A. Zurlo \inst{\ref{inst4} \and \ref{inst16}}
}

\institute{
        INAF -- Osservatorio Astronomico di Padova, Vicolo dell'Osservatorio 5, I-35122, Padova, Italy \label{inst1}
\and Dipartimento di Fisica a Astronomia "G. Galilei", Universita' di Padova, Via Marzolo, 8, 35121 Padova, Italy \label{inst2}
\and LESIA, Observatoire de Paris, PSL Research University, CNRS, Sorbonne Universités, UPMC Univ. Paris 06, Univ. Paris Diderot, Sorbonne Paris Cité \label{inst3}
\and CRAL, UMR 5574, CNRS, Universit\'{e} Lyon 1, 9 avenue Charles Andr\'{e}, 69561 Saint Genis Laval Cedex, France \label{inst21}
\and Aix Marseille Univ, CNRS, LAM, Laboratoire d'Astrophysique de Marseille, Marseille, France \label{inst4}
\and University of Atacama, Copayapu 485, Copiapo, Chile \label{cile}
\and Institute of Astronomy, University of Cambridge, Madingley Road, Cambridge CB3 0HA, UK \label{inst20}
\and Konkoly Observatory, Research Centre for Astronomy and Earth Sciences, Hungarian Academy of Sciences, P.O. Box 67, H-1525 Budapest, Hungary \label{inst19}
\and European Southern Observatory, Karl Schwarzschild St, 2, 85748 Garching, Germany \label{inst17}
\and Max-Planck Institute for Astronomy, K\"onigstuhl 17, 69117 Heidelberg, Germany \label{inst5}
\and Instituto de F\'isica y Astronom\'ia, Facultad de Ciencias, Universidad de Valpara\'iso, Av. Gran Breta\~na 1111, Playa Ancha, Valpara\'iso, Chile \label{inst18}
\and Universit\'e Grenoble Alpes, IPAG, 38000 Grenoble, France \label{inst9}
\and INAF-- Osservatorio Astrofisico di Arcetri, L.go E. Fermi 5, 50125 Firenze, Italy \label{inst30}
\and Institute for Astronomy, ETH Zurich, Wolfgang-Pauli Strasse 27, 8093 Zurich, Switzerland \label{inst6}
\and Universidad de Chile, Camino el Observatorio, 1515 Santiago, Chile \label{inst7}
\and Department of Physics, University of Oxford, Parks Rd, Oxford OX1 3PU, UK \label{inst8}
\and Institute for Astronomy, The University of Edinburgh, Royal Observatory, Blackford Hill, Edinburgh, EH9 3HJ, U.K. \label{inst11}
\and INAF-- Osservatorio Astronomico di Catania, Via Santa Sofia, 78, Catania, Italy  \label{inst40}   
\and INAF-- Osservatorio Astronomico di Capodimonte, Salita Moiariello 16, 80131 Napoli, Italy \label{inst50}
\and Observatoire Astronomique de l'Universit\'e de Gen\`eve, Chemin des Maillettes 51, 1290 Sauverny, Switzerland \label{inst12}
\and Leiden Observatory,Niels Bohrweg 2, NL-2333 CA Leiden, The Netherlands \label{inst13}   
\and Department of Astronomy, Stockholm University, AlbaNova University Center, 106 91 Stockholm, Sweden \label{inst14}
\and INAF-- Istituto di Astrofisica Spaziale e Fisica Cosmica di Milano, via E. Bassini 15, 20133 Milano, Italy \label{inst60}
\and Department of Astronomy, University of Michigan, 1085 S. University, Ann Arbor, MI 48109 \label{inst15}
\and European Southern Observatory (ESO), Alonso de C\'ordova 3107, Vitacura, Casilla 19001, Santiago, Chile \label{inst007}
\and N\'ucleo de Astronom\'ia, Facultad de Ingenier\'ia, Universidad Diego Portales, Av. Ejercito 441, Santiago, Chile \label{inst16}
            }

\abstract {A large number of systems harboring a debris disk show evidence for a double belt architecture.
One hypothesis for explaining the gap between the debris belts in these disks is the presence of one or more planets dynamically carving it. For this reason these disks represent prime targets for searching planets using direct imaging instruments, like VLT/SPHERE.} {The goal of this work is to investigate this scenario in systems harboring debris disks divided into two components, placed, respectively, in the inner and outer parts of the system. All the targets in the sample were observed with the SPHERE instrument, which performs high-contrast direct imaging, during the SHINE GTO. Positions of the inner and outer belts were estimated by SED fitting of the infrared excesses or, when available, from resolved images of the disk. Very few planets have been observed so far in debris disks gaps and we intended to test if such non-detections depend on the observational limits of the present instruments. This aim is achieved by deriving theoretical predictions of masses, eccentricities and semi-major axes of planets able to open the observed gaps and comparing such parameters with detection limits obtained with SPHERE.}{The relation between the gap and the planet is due to the chaotic zone neighboring the orbit of the planet. The radial extent of this zone  depends on the mass ratio between the planet and the star, on the semi-major axis and on the eccentricity of the planet and it can be estimated analytically. We first tested the different analytical predictions using a numerical tool for the detection of chaotic behaviour and then selected the best formula for estimating the planet physical and dynamical properties required to open the observed gap.  We then apply the formalism to the case of one single planet on a circular or eccentric orbit. We then consider multi-planetary systems: two and three equal-mass planets on circular orbits and two equal-mass planets on eccentric orbits in a packed configuration. As a final step, we compare each couple of values ($M_p$,$a_p$), derived from the dynamical analysis of single and multiple planetary models, with the  detection limits obtained with SPHERE. } {For one single planet on a circular orbit we obtain conclusive results that allow us to exclude such hypothesis since in most cases this configuration requires massive planets which should have been detected by our observations. Unsatisfactory is also the case of one single planet on an eccentric orbit for which we obtained  high masses and/or eccentricities which are still at odds with observations. Introducing multi planetary architectures is encouraging because for the case of three packed equal-mass planets on circular orbits we obtain quite low masses for the perturbing planets which would remain undetected by our SPHERE observations. Also the case of two equal-mass planets on eccentric orbits is of interest since it suggests the possible presence of planets with masses lower than the detection limits and with moderate eccentricity. Our results show that the apparent lack of planets in gaps between double belts could be explained by the presence of a system of two or more planets possibly of low mass and on an eccentric orbits whose sizes are below the present detection limits. }{}
\keywords{Methods: analytical, data analysis, observational - Techniques: high angular resolution, image processing - Planetary systems - Kuiper belt - Planet-disk interactions}

\maketitle

\section{Introduction}
Debris disks are optically thin, almost gas-free dusty disks observed around a significant fraction of main-sequence stars \citep[20-30$\%$, depending on the spectral type, see][]{Matthews1} older than about $10$ Myr. Since the circumstellar dust is short-lived, the very existence of these disks is considered as an evidence that dust-producing planetesimals are still present in mature systems, in which planets have formed- or failed to form a long time ago \citep{Krivov,Moro, Wyatt}. It is inferred that these planetesimals orbit their host star from a few to tens or hundreds of AU, similarly to the Asteroid ($\sim 2.5$ AU) and Kuiper belts ($\sim 30$ AU), continually supplying fresh dust through mutual collisions. \\
Systems that harbor debris disks have been previously investigated with high-contrast imaging instruments in order to infer a correlation between the presence of planets and second generation disks. The first study in this direction was performed by \cite{Apai} focused on the search of massive giant planets in the inner cavities of 8 debris disks with VLT/NACO. Additional works followed like, for example, the NICI \citep{Wahhaj1}, the SEEDS \citep{Janson} statistical study of planets in systems with debris disks and the more recent VLT/NaCo survey performed with the Apodizing Phase Plate on six systems with holey debris disks \citep{Meshkat}. However, in all these studies single-component and multi-components debris disks 
were mixed up and the authors did not perform a systematic analysis on putative planetary architectures possibly matching observations. Similar and more detailed analysis can be found for young and nearby stars with massive debris disks such as Vega, Fomalhaut and $\epsilon$ Eri \citep{Janson1} or $\beta$ Pic \citep{Lagrange1}.\\
In this context, the main aim of this work was to analyze systems harboring a debris disk composed of two belts, somewhat similar to our Solar System: a warm Asteroid-like belt in the inner part of the system and a cold Kuiper-like belt farther out from the star. The gap between the two belts is assumed to be almost free from planetesimals and grains. In order to explain the existence of this empty space, the most straightforward assumption is to assume the presence of one or more planets orbiting the star between the two belts \citep{Kanagawa, Su1, Kennedy, Schuppler, Shannon}.\\
The hypothesis of a devoid gap may not always be correct. Indeed, dust grains which populate such region may be too faint to be detectable with spatially resolved images and/or from SED fitting of infrared excesses. This is for example the case for HD131835, that seems to have a very faint component (barely visible from SPHERE images) between the two main belts \citep{Feldt}. In such cases the hypothesis of no free dynamical space between planets that we will adopt in Section 6 could be relaxed and smaller planets, very close to the inner and outer belts respectively, could be the responsible for the disk's architecture. However, these kind of scenarios introduce degeneracies that cannot be validated with current observation whereas a dynamically full system is described by a univocal set of parameters that we can promptly compare with our data. \\ 
In this paper we follow the assumption of planets as responsible for gaps which appears to be the most simple and appealing interpretation for double belts. We explore different configurations in which one or more planets are evolving on stable orbits within the two belts with separations which are just above their stability limit (packed planetary systems) and test the implications of adopting  different values of mass and orbital eccentricity. We acknowledge that other dynamical  mechanisms may be at play possibly leading to more complex scenarios characterized by further rearrangement of the planets architecture. Even if planet--planet scattering in most cases cause the disruption of the planetesimal belts during the chaotic phase \citep{Marzari1}, more gentle evolutions may occur like that invoked for the solar system. In this case, as described by \cite{Levinson}, the scattering of Neptune by Jupiter and the subsequent outward migration of the outer planet by planetesimal scattering lead to a configuration in which the planets have larger separations compared to that predicted only from dynamical stability. Even if we do not contemplate these more complex systems, our model gives an idea of the
minimum requirements in terms of mass and orbital eccentricity for a system of planets to carve the observed double--belts. Even more exotic scenarios may be envisioned in which  a planet near to the observed belt would have scattered inward a large  planetesimal which would have subsequently impacted a planet causing the formation of a great amount of dust \citep{Kral1}. However, according to \cite{Geiler}, in the vast majority of debris disks, which include also many of the systems analyzed in this paper, the warm infrared excess is compatible with a natural dynamical evolution of a primordial asteroid-like belt (see Section 2). The possibility that a recent energetic event is responsible for the inner belt appears to be remote as a general explanation for the stars in our sample. \\ 
One of the most famous systems with double debris belts is HR8799 \citep{Su2}.  Around the central star and in the gap between the belts, four giant planets were observed each of which has a mass in the range $[5,10]$ $M_J$ \citep{Marois2, Marois1, Zurlo} and there is room also for a fifth planet \citep{Booth2}. This system suggests that multi-planetary and packed architectures may be  common in extrasolar systems.\\
Another interesting system is HD95086 that harbors a debris disk divided into two components and has a known planets that orbits between the belts. The planet has a mass of $\sim5$ $M_J$ and was detected at a distance of $\sim 56$ AU \citep{Rameau}, close to the inner edge of the outer belt at $61$ AU. Since the distance between the belts is quite large, the detected planet is unlikely to be the only responsible for the entire gap and multi-planetary architecture may be invoked \citep{Su1}.\\ 
HR8799, HD95086 and other similar systems seem to point to some correlations between planets and double components disks and a more systematic study of such systems is the main goal of this paper.
However, up to now very few giant planets were found orbiting far from their stars, even with the help of the most powerful direct imaging instruments such as SPHERE or GPI \citep{Bowler}. For this reason, in the hypothesis, that  the presence of one or more  planets are responsible for the gap in double debris belts systems, we have to estimate
the dynamical and physical properties of these potentially undetected objects.\\  
We analyze in the following a sample of systems having debis disks with two distinct components 
determined  by fitting the spectral energy distribution (SED) from \emph{Spitzer Telescope} data (coupled with previous flux measurements) and observed also with the SPHERE instrument that performs high contrast direct imaging searching for giant exoplanets. The two belts architecture and their radial location obtained by \cite{Chen} was also confirmed in the vast majority of cases by \cite{Geiler} in a further analysis.
In our analysis of these double belts we assume that the gap between the two belts is due to the presence of one, two or three planets in circular or eccentric orbits. In each case we compare the model predictions in terms of masses, eccentricities and semi-major axis of the planets with the SPHERE instrument detection limits to test their
observability. In this way we can put stringent constraints on the potential planetary system responsible for each double belt.
This kind of study was already applied at HIP67497 and published in \cite{Bonnefoy}.\\
Further analysis involves the time needed for planets to dig the gap. In \cite{Shannon} they find a relation between the typical time scales, $t_{clear}$, for the creation of the gap and the numbers of planets, $N$, between the belts as well as their masses: for a given system's age they can thus obtain the minimum masses of planets that could have carved out the gap as well as their typical number. Such information is particularly interesting for young systems because in these cases we can put a lower limit on the number of planets that orbit in the gap. We will not take into account this aspect in this paper since our main purpose is to present a dynamical method but we will include it in further statistical studies. Other studies, like the one published by \cite{Nesvold1}, directly link the width of the dust-devoid zone (chaotic zone) around the orbit of the planet with the age of the system. From such analysis emerges that the resulting gap for a given planet may be wider than expected from classical calculations of the chaotic zone \citep{Wisdom, Mustill2}, or, equivalently, observed gaps may be carved by smaller planets. However, these results are valid only for ages $\le10^{7}$ yrs and, since very few systems in this paper are that young, we will not take into account the time dependence in chaotic zones equations.\\  
In Section 2 we illustrate how the targets were chosen; in Section 3 we characterize the edges of the inner and outer components of the disks; in Section 4 we describe the technical characteristics of the SPHERE instrument and the observations and data reduction procedures; in Section 5 and 6 we present the analysis performed for the case of one planet, and two or three planets, respectively; in Section 7 we study more deeply some individual systems of the sample; in Section 8, finally, we provide our conclusions.

\section{Selection of the targets}
 \begin{table*}
\caption{Stellar parameters for directly imaged systems with SPHERE. For each star we show spectral type, mass (in solar mass units), luminosity (in solar luminosity units), age and distance.}
\label{tabu1}   
\centering
\begin{threeparttable}
\begin{tabular}{c c c c c c}
\hline\hline
Name &Spec Type & $M_*/M_{\odot}$ &$L_*/L_{\odot}$&Age & Dist \\
&&&&(Myr)&(pc)\\
\hline
\\                     
HD1466   & F8&1.1 &1.6& 45$^{+5}_{-10}$ &42.9$ \pm0.4 $ \tnote{a} \\[0.75ex]
HD3003  & A0&2.1 &18.2& 45$^{+5}_{-10}$ &45.5$ \pm 0.4$ \tnote{b}   \\[0.75ex]
HD15115  &F2 & 1.3&3.9& 45$^{+5}_{-10}$ & 48.2$ \pm1 $ \tnote{a}\\[0.75ex]
HD30447&F3 & 1.3&3.9& 42$^{+8}_{-7}$ & 80.3$ \pm1.6 $ \tnote{a}  \\[0.75ex]
HD35114&F6&1.2&2.3&42$^{+8}_{-7}$ &47.4$ \pm 0.5 $ \tnote{a}\\[0.75ex]
$\zeta$ Lep &A2&1.9&21.6&300$\pm$180  &21.6$ \pm 0.1 $ \tnote{b}\\[0.75ex]
HD43989 & F9& 1.1&1.6&  45$^{+5}_{-10}$ & 51.2$ \pm 0.8 $ \tnote{a} \\[0.75ex]
HD61005 &G8 & 0.9&0.7&50$^{+20}_{-10}$ &36.7$ \pm 0.4 $ \tnote{a}   \\[0.75ex]
HD71155 &A0 & 2.4& 40.5&260$\pm$75 & 37.5$ \pm 0.3 $ \tnote{b} \\[0.75ex]
HD75416 &B8 & 3&106.5& 11$\pm$3 & 95$ \pm 1.4$ \tnote{b} \\[0.75ex]
HD84075 &G2 & 1.1&1.4& 50$^{+20}_{-10}$&62.9 $ \pm 0.9  $ \tnote{a}  \\[0.75ex]
HD95086 &A8 & 1.6&8& 16$\pm$5& 83.8$ \pm 1.9 $ \tnote{a} \\[0.75ex]
$\beta$ Leo  & A3& 1.9&14.5&50$^{+20}_{-10}$ & 11$ \pm 0.1$ \tnote{b}  \\[0.75ex]
HD106906 & F5& 2.7 \tnote{c}&6.8&16$\pm$5 & 102.8$ \pm 2.5 $ \tnote{a} \\[0.75ex]
HD107301 &B9 & 2.4&42.6&16$\pm$5  &93.9$ \pm 3$ \tnote{b}  \\[0.75ex]
HR4796   & A0& 2.3&26.8&  10$\pm$3 & 72.8$ \pm 1.7 $ \tnote{b}\\[0.75ex]
$\rho$ Vir &A0&1.9&15.9&100$\pm$80&36.3$ \pm 0.3 $ \tnote{b} \\[0.75ex]
HD122705 &A2 & 1.8& 12.3&17$\pm$5 &112.7$ \pm 9.3$ \tnote{b} \\[0.75ex]
HD131835 &A2 & 1.9&15.8& 17$\pm$5 & 145.6$ \pm 8.5$ \tnote{a} \\[0.75ex]
HD133803 &A9 & 1.6&6.2&17$\pm$5  & 111.8$ \pm 3.3 $ \tnote{a} \\[0.75ex]
$\beta$ Cir &A3 & 2&18.5&400$\pm$140 &30.6$ \pm 0.2$ \tnote{b} \\[0.75ex]
HD140840 &B9 & 2.3&37.4&17$\pm$5 &165$ \pm 10.4$ \tnote{a} \\[0.75ex]
HD141378&A5 & 1.9& 17& 380$\pm$190 & 55.6$ \pm2.1 $ \tnote{a}\\[0.75ex]
$\pi$ Ara &A5&1&13.7&600$\pm$220   &44.6$ \pm 0.5$ \tnote{b}\\[0.75ex]
HD174429 &G9 & 1&1.6&24$\pm$5  &51.5$ \pm 2.6$ \tnote{b} \\[0.75ex]
HD178253 &A2 & 2.2& 31&380$\pm$90    & 38.4$ \pm 0.4 $ \tnote{b}\\[0.75ex]
$\eta$ Tel      & A0&2.2 & 21.3& 24$\pm$5  & 48.2$ \pm 0.5$ \tnote{b} \\[0.75ex]
HD181327 & F6& 1.3&3& 24$\pm$5 & 48.6$ \pm 1.1 $ \tnote{a} \\[0.75ex]
HD188228 &A0 & 2.3&26.6& 50$^{+20}_{-10}$ &32.2$ \pm 0.2 $ \tnote{b} \\[0.75ex]
$\rho$ Aql & A2& 2.1&21.6& 350$\pm$150& 46$ \pm 0.5  $ \tnote{b} \\[0.75ex]
HD202917 &G7 & 0.9& 0.8& 45$^{+5}_{-10}$ & 47.6$ \pm 0.5 $ \tnote{a}\\[0.75ex]
HD206893 &F5&1.3&3& 250$^{+450}_{-200}$ &40.7$ \pm 0.4 $ \tnote{a}\\[0.75ex]
HR8799  &A5 & 1.5&8& 42$^{+8}_{-7}$ & 40.4$ \pm 1 $ \tnote{a} \\[0.75ex]
HD219482 &F6 & 1&2.3&400$^{+200}_{-150}$ &20.5$ \pm 0.1$ \tnote{b} \\[0.75ex]
HD220825 &A0 & 2.1&22.9& 149$^{+31}_{-49}$ &47.1$ \pm 0.6 $ \tnote{b} \\[0.75ex]

\hline

\end{tabular}
\begin{tablenotes}
\item[a] Gaia parallaxes \citep{Lindegren}; \item[b] Hipparcos parallaxes \citep{vanLeeuwen}; \item[c] Mass of the star from \cite{Bonavita}.
\end{tablenotes}
\end{threeparttable}
\end{table*}

The first step of this work is to choose the targets of interest. For this purpose, we use the  published catalog of \cite{Chen} (from now on C14) in which they have calibrated the spectra of $571$ stars looking for excesses in the infrared from $5.5$ $\mu m$ to $35$ $\mu m$ (from the \textit{Spitzer}  survey) and when available (for $473$ systems) they also used the MIPS $24$ $\mu m$ and/or $70$ $\mu m$ photometry to calibrate and better constrain the SEDs of each target. These systems cover a wide range of spectral type (from B9 to K5, corresponding to stellar masses from $0.5$ $M_{\odot}$ to $5.5$ $M_{\odot}$) and ages (from $10$ Myr to $1$ Gyr) with the majority of targets within $200$ pc from the Sun.\\
In \cite{Chen}, the fluxes for all $571$ sources were measured in two bands, one at $8.5-13$ $\mu m$ to search for weak $10$ $\mu m$ silicate emission and another at $30-34$ $\mu m$ to search for long wavelengths excess of cold grains. Then, the excesses of SEDs were modeled  using zero, one and two blackbodies because debris disks spectra typically do not have strong spectral features and  blackbody modeling provides typical dust temperatures. We select amongst the entire sample only systems with two distinct black-body temperatures, as obtained by \cite{Chen}.\\
SED fitting alone suffers from degeneracies and in some cases systems classified as double belts can also be fitted as singles belt changing belt width, grains properties, etc.
However, in \cite{Geiler}, the sample of 333 double belt systems of C14 was reanalyzed to investigate the effective presence of an inner component. In order to perform their analysis, they excluded 108 systems for different reasons (systems with temperature of black-body $T_{1,BB}\le30K$ and/or $T_{2,BB}\sim500K$; systems for which one of the two components was too faint with respect to the other or for which the two components had similar temperatures; systems for which the fractional luminosity of the cold component was less than $4 \times 10^{-6}$) and they ended up with 225 systems that they considered to be reliable two-component disks. They concluded that of these 225 stars, 220 are compatible with the hypothesis of a two-component disk, thus $98\%$ of the objects of their sample. Furthermore, they pointed out that the warm infrared excesses for the great majority of the systems are compatible with a natural dynamical evolution of inner primordial belts. The remaining $2\%$ of the warm excesses are too luminous and may be created by other mechanisms, for example  by transport of dust grains from the outer belt to the inner regions \citep{Kral}.\\
The 108 discarded systems are not listed in \cite{Geiler} and we cannot fully crosscheck all of them with the ones in our sample. However, two out of three exclusion criteria (temperature and fractional luminosity) can be replicated just using parameters obtained by \cite{Chen}. This results in 79 objects not suitable for their analysis and between them only HD71155, $\beta$ Leo and HD188228 are in our sample. We cannot crosscheck the last 29 discarded systems but, since they are a minority of objects, we can apply our analysis with good confidence.\\  
Since the gap between belts typically lies at tens of AU from the central star the most suitable planet hunting technique to detect planets in this area is direct imaging. Thus, we crosschecked this restricted selection of objects of the C14 with the list of targets of the SHINE GTO survey observed with SPHERE up to February $2017$ (see Section 4). A couple of targets with unconfirmed candidates within the belts are removed from the sample as the interpretation of these systems will heavily depend on the status of these objects.
We end up with a sample of $35$ main-sequence young stars ($t\le600$ Myr) in a wide range of spectral types, within $150$ pc from the Sun. Stellar properties are listed in Table \ref{tabu1}.  We adopted Gaia parallaxes \citep{Lindegren}, when available, or Hipparcos distances as derived by \cite{vanLeeuwen}. Masses were taken from C14 (no error given) while luminosities have been scaled to be consistent with the adopted distances. The only exception is HD106906 that was discovered to be a close binary system after the C14 publication and for which we used the mass as given by Lagrange et al., submitted. Ages, instead, were obtained using the method described in Section 4.2. \\

\section{Characterization of gaps in the disks}
 For each of the systems listed in Table \ref{tabu2}, temperatures of the grains in the two belts, $T_{1,BB}$ and $T_{2,BB}$, were available from C14. Then, we obtain the blackbody radii of the two belts \citep{Wyatt} using the equations 
 \begin{equation}
 R_{i,BB}=\Biggl (\frac{278K}{T_{i,BB}}\Biggr)^2\Biggl(\frac{L_*}{L_{\odot}}\Biggr)^{0.5} AU,
 \end{equation} 
 with $i=1,2$.\\
 However, if the grains do not behave like perfect blackbodies a third component, the size of dust particles, must be taken into account. Indeed, now the same SED could be produced by smaller grains further out or larger particles closer to the star. Therefore, in order to break this degeneracy, we searched in literature for debris disks in our sample that have been previously spatially resolved using direct imaging. In fact, from direct imaging data many peculiar features are clearly visible and sculptured edges are often well constrained. We found $19$ resolved objects and used positions of the edges as given by images of the disks (see Appendix B). We note, however, that usually disks resolved at longer wavelengths are much less constrained than the ones with images in the near IR or in scattered light and only estimations of the positions of the edges are possible. Moreover, images obtained in near IR and visible wavelengths usually have higher angular resolutions. This is not the case for ALMA that works at sub-millimeter wavelengths using interferometric measurements with resulting high angular resolutions. For these reasons, we preferred for a given system images of the disks at shorter wavelengths and/or, when available, ALMA's data. \\
 For all the other undetected disks by direct imaging, we applied the correction factor $\Gamma$ to the black-body radius of the outer belt which depends on a power law of the luminosity of the star expressed in solar luminosity \citep{Pawellek2}. More details on the $\Gamma$ coefficient are given in Appendix A.  
We indicate with $R_{2}$ in Table \ref{tabu2} the more reliable disk radii obtained multiplying $R_{2,BB}$ by  $\Gamma$.\\
The new radii that we obtain need a further correction to be suitable for our purposes. Indeed, they refer to the mid-radius of the planetesimal belt, since we predict that the greater part of the dust is produced there, and do not represent the inner edge of the disk that is what we are looking for in our analysis. Such error should be greater with increasing distance of the belt from the star. Indeed, we expect that the farther the disk is placed the wider it is due to the weaker influence of the central star and to effects like Poynting-Robertson and radiation pressure \citep{Krivov, Moro}. Thus, starting from data of resolved disks (see Table \ref{tableA}) we adopt a typical value of $\Delta R/R_2=0.2$, where $\Delta R= R_2-d_2$ is the difference between the estimated position of the outer belt and the inner edge of the disk. Such value for the relative width of the outer disk is also supported by other systems that are not in our sample as, for example, $\epsilon$ Eridani for which $\Delta R=0.17$ \citep{Booth3} or as the Solar System itself for which $\Delta R=0.18$ for the Kuiper belt \citep{Gladman1}. Estimated corrected radii and inner edges for each disk can be found in Table \ref{tabu2} in columns $6$ ($R_2$) and $7$ ($d_2$).\\
Following the same arguments of \cite{Geiler}, we do not apply the $\Gamma$ correction to the inner component since it differs significantly with respect to the outer one having, for example, quite different sizes distribution and composition of grains. However, for the inner belts results from black-body analysis should be more precise as confirmed in the few cases in which the inner component was resolved \citep{Moerchen}. Moreover, in systems with radial velocity planets we can apply the same dynamical analysis that we present in this paper and estimate the position of the inner belt. Indeed, for RV planets the semi-major axis, the eccentricity and the mass (with an uncertainty of $\sin{i}$) are known and we can estimate the width of the clearing zone and compare it with the expected position of the belt. Results from such studies seem to point to a correct placement of the inner component from SED fitting (Lazzoni et al., in prep).\\
In Table \ref{tabu2} we show for each system in the sample the temperature of the inner and the outer belts, $T_{1,BB}$ and $T_{2,BB}$, the black-body radius of the inner and outer belts, $R_{1,BB}$ and $R_{2,BB}$ respectively. In column $d_{2,sol}$ of Table \ref{tabu2} we show the positions of the inner edge as given by direct imaging data for spatially resolved systems. We also want to underline that the systems in our sample are resolved only in their farther component, with the exception of HD71155 and $\zeta$ Lep that also have resolved inner belts \citep{Moerchen}. Indeed, the inner belts are typically very close to the star so that for the instruments used it was not possible to separate their contributions from the flux of the stars themselves. We illustrate all the characteristics of the resolved disks in Table \ref{tableA}.\\

\begin{table*} [h!]
\caption{Debris disks parameters for direct imaged systems with SPHERE. $T_{gr,1}$ and $T_{gr,2}$ are the black-body temperatures, $R_{1,BB}$ and $R_{2,BB}$ the black-body radii; $R_2$ and $d_2$ are the "real" radius and inner edge obtained with the $\Gamma$ correction; $d_{2,sol}$ is the position of the inner edge for the spatially resolved systems.  }
\label{tabu2} 
\centering
\begin{threeparttable}
\begin{tabular}{ccccccccc}
 \hline\hline
Name&  $T_{1,BB}$  & $R_{1,BB}$  & $T_{2,BB}$ &  $R_{2,BB}$ &$R_2$& $d_2$ &  $d_{2,sol}$\\
&(K)&(AU)&(K)&(AU)&(AU)&(AU)&(AU)\\
 \hline
 \\[0.5ex]
HD1466  \tnote{b}    &374$^{+7}_{-5}$ & 0.70$\pm0.01$&97$^{+5}_{-7}$ &10.5$^{+0.5}_{-0.8}$ &51.8$^{+2.7}_{-3.7}$ &41.4$^{+2.1}_{-3.0}$ &(...)\\[1ex]
HD3003 \tnote{b}    &472$^{+7}_{-5}$ & 1.50$\pm0.02$&173$^{+5}_{-5}$ & 11.0$\pm0.3$&20.7$\pm0.6$ &16.6$\pm0.5$&(...)\\[1ex]
HD15115 \tnote{a} &182$^{+4}_{-7}$ & 4.6$^{+0.1}_{-0.2}$&54$\pm5$ &52.6$\pm4.9$ &182.4$\pm16.9$ & 145.9$\pm13.5$ &48\\[1ex]
HD30447 \tnote{a}&106$^{+6}_{-5}$ &13.6$^{+0.8}_{-0.6}$ & 57$^{+4}_{-6}$& 47.0$^{+3.3}_{-5.0}$& 163.6$^{+11.5}_{-17.2}$& 130.9$^{+9.2}_{-13.8}$  &60\\[0.75ex]
HD35114 \tnote{b}&139$^{+12}_{-7}$&6.0$^{+0.5}_{-0.3}$&66$^{+10}_{-15}$&26.7$^{+4.0}_{-6.1}$&115.5$^{+17.5}_{-26.3}$&92.4$^{+14.0}_{-21.0}$&(...)\\[0.75ex]
$\zeta$ Lep \tnote{b}&368$\pm5$&2.70$\pm0.04$&133$\pm5$&20.3$\pm0.8$&35.6$\pm1.3$&28.5$\pm1.1$&(...)\\[0.75ex]
HD43989 \tnote{b}& 319$^{+30}_{-26}$ & 1.0$\pm0.1$&66$^{+9}_{-10}$ &22.3$^{+3.0}_{-3.4}$ &111.5$^{+15.2}_{-16.9}$&89.2$^{+12.2}_{-13.5}$ &(...)\\[0.75ex]
HD61005 \tnote{a} & 78$^{+6}_{-4}$& 10.5$^{+0.8}_{-0.5}$& 48$\pm5$&27.6$\pm2.9$ &(...)&(...) &71\\[0.75ex]
HD71155 \tnote{a} & 499$^{+0}_{-7}$& 2$^{+0.00}_{-0.03}$& 109$\pm5$&41.4$\pm1.9$ &56.5$\pm2.6$&45.2$\pm2.1$&69\\[0.75ex]
HD75416 \tnote{b} &393$^{+4}_{-6}$ & 5.2$\pm0.1$&124$^{+7}_{-5}$ &51.9$^{+2.9}_{-2.1}$&48.1$^{+2.7}_{-1.9}$&38.5$^{+2.2}_{-1.6}$&(...)\\[0.75ex]
HD84075 \tnote{b}&149 $^{+23}_{-18}$& 4.1$^{+0.6}_{-0.5}$&54$^{+7}_{-10}$ &31.3$^{+4.1}_{-5.8}$ &164.4$^{+21.3}_{-30.4}$&131.5$^{+17.0}_{-24.4}$ &(...) \\[0.75ex]
HD95086 \tnote{a} &225$^{+10}_{-7}$& 4.3$^{+0.2}_{-0.1}$&57$\pm5$ &67.1$\pm5.9$ &175.6$\pm15.4$&140.5$\pm12.3$ &61 \\[0.75ex]
$\beta$  Leo \tnote{a}  &499$^{+0}_{-9}$ & 1.2$^{+0.00}_{-0.02}$&106$\pm5$ & 26.2$\pm1.2$&53.9$\pm2.5$&43.1$\pm2.0$ &15\\[0.75ex]
HD106906 \tnote{a} &124$^{+11}_{-8}$ & 13.1$^{+1.2}_{-0.8}$&81$^{+7}_{-12}$ &30.7$^{+2.6}_{-4.5}$ &85.6$^{+7.4}_{-12.7}$&68.5$^{+5.9}_{-10.1}$ &56\\[0.75ex]
HD107301 \tnote{b}&246$\pm5$ & 8.3$\pm0.2$& 127$\pm5$&31.3$\pm1.2$&41.8$\pm1.6$& 33.5$\pm1.3$& (...)\\[0.75ex]
HR4796 \tnote{a}  &231$^{+5}_{-6}$ & 7.5$\pm0.2$& 97$\pm5$&42.6$\pm2.2$ &68.5$\pm3.5$& 54.8$\pm2.8$&73 \\[0.75ex]
$\rho$ Vir \tnote{a}&445$^{+6}_{-7}$&1.60$\pm0.02$&78$\pm5$&50.6$\pm3.2$&100.5$\pm6.4$&80.4$\pm5.2$&98\\[0.75ex]
HD122705 \tnote{b}&387$^{+5}_{-6}$& 1.80$^{+0.02}_{-0.03}$&127$^{+6}_{-8}$ & 16.8$^{+0.8}_{-1.1}$&36.9$^{+1.7}_{-2.3}$&29.6$^{+1.4}_{-1.9}$ &(...)\\[0.75ex]
HD131835 \tnote{a} & 216$\pm5$& 6.6$\pm0.2$&78$\pm5$ & 50.5$\pm3.2$& 100.4 $\pm6.4$&80.4$\pm5.2$  &89\\[0.75ex]
HD133803 \tnote{b}& 368$\pm5$& 1.40$\pm0.02$& 142$\pm5$&9.6$\pm0.3$ &27.6$\pm1.0$&22.1$\pm0.8$  &(...)\\[0.75ex]
$\beta$ Cir \tnote{b}&387$^{+6}_{-7}$ & 2.20$^{+0.03}_{-0.04}$& 155$^{+5}_{-7}$&13.8$^{+0.4}_{-0.6}$ &25.8 $^{+0.8}_{-1.2}$&20.7$^{+0.7}_{-0.9}$&(...)\\[0.75ex]
HD140840 \tnote{b}& 341$^{+4}_{-7}$&4.1$\pm0.1$ &88$\pm5$ &61.0$\pm3.5$ & 86$\pm4.9$ & 68.8$\pm3.9$ &(...)\\[0.75ex]
HD141378 \tnote{a} & 347$^{+7}_{-5}$& 2.60$^{+0.05}_{-0.04}$&69$\pm5$ &66.9$\pm4.8$ &129.3$\pm9.4$& 103.4$\pm7.5$ &133\\[0.75ex]
$\pi$ Ara \tnote{a}&173$\pm5$&9.6$\pm0.3$&54$^{+6}_{-4}$&98.2$^{+10.9}_{-7.3}$&206.7$^{+23.0}_{-15.3}$&165.3$^{+18.4}_{-12.2}$&122\\[0.75ex]
HD174429 \tnote{b}&460$^{+39}_{-67}$ &0.50$^{+0.04}_{-0.07}$ & 39$\pm7$& 64.5$\pm11.6$&319.8$\pm57.4$&255.8$\pm45.9$ &(...) \\[0.75ex]
HD178253 \tnote{b}&307$\pm6$ & 4.6$\pm0.1$&100$^{+5}_{-7}$ &43.0$^{+2.2}_{-3.0}$ &65.4$^{+3.3}_{-4.6}$& 52.3$^{+2.6}_{-3.7}$&(...)\\[0.75ex]
$\eta$ Tel \tnote{a}      &  277$^{+5}_{-9}$& 4.7$^{+0.1}_{-0.2}$&115$^{+4}_{-7}$ & 27.0$^{+0.9}_{-1.6}$&47.6$^{+1.7}_{-2.9}$ &38.1$^{+1.3}_{-2.3}$ &24\\[0.75ex]
HD181327 \tnote{a}& 94$\pm5$& 15.3$\pm0.8$&60$\pm5$ & 37.5 $\pm3.1$&144$\pm12$&115.2$\pm9.6$ &70\\[0.75ex]
HD188228 \tnote{a} & 185$^{+37}_{-56}$& 11.6$^{+2.3}_{-3.5}$& 72$\pm6$&76.9$\pm6.4$ &124.2$\pm10.3$&99.3$\pm8.3$ &107\\[0.75ex]
$\rho$ Aql \tnote{a} &268$^{+6}_{-5}$ &5.0$\pm0.1$ &66$\pm5$ &82.4$\pm6.2$ &144.7$\pm11.0$& 115.8$\pm8.8$&223\\[0.75ex]
HD202917 \tnote{a}& 289$^{+47}_{-33}$& 0.8$\pm0.1$& 75$^{+5}_{-6}$& 11.9$^{+0.8}_{-1.0}$&(...) &(...)  &61\\[0.75ex]
HD206893 \tnote{a}&499$^{+0}_{-10}$&0.50$^{+0.00}_{-0.01}$&48$\pm5$&57.7$\pm6.0$&224.3$\pm23.4$&179.4$\pm18.7$&53\\[0.75ex]
HR8799 \tnote{a} &155$^{+6}_{-8}$ & 9.1$^{+0.4}_{-0.5}$ &33$^{+5}_{-3}$ & 200.7$^{+30.4}_{-18.2}$&524.2$^{+79.4}_{-47.7}$&419.4$^{+63.5}_{-38.1}$&101\\[0.75ex]
HD219482 \tnote{b}&423$^{+11}_{-8}$ & 0.70$^{+0.02}_{-0.01}$&78$\pm5$ & 19.2$\pm1.2$&82.8$\pm5.3$&66.2$\pm4.2$&(...) \\[0.75ex]
HD220825 \tnote{b}& 338$^{+8}_{-9}$& 3.2$\pm0.1$& 170$\pm7$& 12.8$\pm0.5$&21.9$\pm0.9$ &17.6$\pm0.7$&(...)\\[0.75ex]
 \\[0.5ex]
\hline
\end{tabular}
\begin{tablenotes}
\item[a] spatially resolved system, used position of the farther edge $d_{2,sol}$ ; \item[b] spatially unresolved system, used position of  the farther edge $d_{2}$ ; 
\end{tablenotes}
\end{threeparttable}
\end{table*}

\section{SPHERE observations}

\subsection{Observations and data reduction}
The Spectro-Polarimetric High-contrast Exoplanet REsearch instrument is installed at the VLT \citep{Beuzit} and is designed to perform high-contrast imaging and spectroscopy in order to find giant exoplanets around relatively young and bright stars.  It is equipped with an extreme adaptive optics system, SAXO \citep{Fusco, Petit}, using a 41 x 41 actuators, pupil stabilization, differential tip-tilt control. The SPHERE instrument has several coronagraphic devices for stellar diffraction suppression, including apodized pupil Lyot coronagraphs \citep{Carbillet} and achromatic four-quadrants phase masks \citep{Boccaletti}. 
The instrument has three science subsystems: the Infra-Red Dual-band Imager and Spectrograph \citep[IRDIS,][]{Dohlen}, an Integral Field Spectrograph \citep[IFS,][]{Claudi} and the Zimpol rapid-switching imaging polarimeter \citep[ZIMPOL,][]{Thalmann}.
Most of the stars in our sample were observed in IRDIFS mode with IFS in the $YJ$ mode and IRDIS in dual-band imaging mode \citep[DBI;][]{Vigan1} using the $H2H3$ filters. Only HR8799, HD95086 and HD106906 were also observed in a different mode using IFS in the YH mode and IRDIS with K1K2 filters.\\
Observations settings are listed for each system in Table \ref{obs}.\\
Both IRDIS and IFS data were reduced at the SPHERE data center hosted at OSUG/IPAG in Grenoble using the SPHERE Data Reduction Handling (DRH) pipeline \citep{Pavlov} complemented by additional dedicated procedures for IFS \citep{Mesa} and the dedicated Specal data reduction software (Galicher in prep.) making use of high contrast algorithms such as PCA, TLOCI, CADI.\\
Further details and references can be found in the various papers presenting SHINE (SpHere INfrared survEy) results on individual targets, e.g., \cite{Samland}.
The observations and data analysis procedures of the SHINE survey will be fully described in Langlois et al. (2017, in prep), along with the
companion candidates and their classification for the data acquired up  to now.\\
Some of the datasets considered in this study  were previously published in \cite{Maire}, \cite{Zurlo}, \cite{Lagrange}, \cite{Feldt}, \cite{Milli2}, \cite{Olofsson} and one is from papers already submitted (HIP 107412: Delorme et al.).

\begin{table*} [h!]
\caption{Observations of the sample. DIT ((detector integration time) refers to the single exposure time, NDIT (Number of Detector InTegrations) to the number of frames in a single data cube.}
\label{obs}
\centering
\begin{tabular}{cccccccc}
\hline\hline
Name & Date& \multicolumn{2}{c}{IFS}&\multicolumn{2}{c}{IRDIS}&Angle ($^{\circ}$)&Seeing ('')\\
           &        & NDIT&DIT&NDIT&DIT&&\\
\hline  
HD1466      &  2016-09-17   &      80   &   64    &   80   &    64   & 31.7   &   0.93 \\
HD3003      &  2016-09-15   &      80   &   64    &  80    &     64  & 32.2   &   0.52\\
HD15115    &  2015-10-25   &      64   &   64     &  64   &     64   & 23.7  &   1.10\\
HD30447    &  2015-12-28   &      96   &   64    &   96   &     64  & 19.5   &   1.36\\ 
HD35114    &  2017-02-11   &       80   &   64    &   80   &     64   & 72.3   &   0.81\\
$\zeta$ Lep&  2017-02-10   &     288   &   16    &  288  &    16   & 80.8   &   0.74\\
HD43989    &  2015-10-27   &      80   &    64    &   80   &    64  &  49.0   &   1.00\\
HD61005    &  2015-02-03   &      64   &    64    &   64   &    64  &  89.3   &   0.50\\
HD71155    &  2016-01-18   &    256    &   16    &  512  &      8  &  36.3   &   1.49\\
HD75416    &  2016-01-01   &      80    &   64    &   80   &    64  &  22.3   &   0.89\\
HD84075    &  2017-02-10   &      80    &   64    &   80   &    64  &  24.6   &    0.43\\
HD95086    &  2015-05-11   &      64    &   64    &  256  &    16  &   23.0   &   1.15\\
$\beta$ Leo&  2015-05-30   &    750    &    4     &  750  &      4  &   34.0   &   0.91\\
HD106906  &  2015-05-08   &      64    &   64    &  256  &    16  &   42.8   &   1.14\\
HD107301  &  2015-06-05   &      64    &   64    &    64  &    64  &   23.8   &   1.36\\
HR4796      &  2015-02-03   &      56    &   64    &  112  &    32  &   48.9   &   0.57\\
$\rho$ Vir    &  2016-06-10   &      72    &   64    &   72   &    64  &   26.8  &    0.66\\
HD122705   &  2015-03-31   &     64    &   64    &   64   &    64  &   35.9   &    0.87\\
HD131835   &  2015-05-14   &     64    &   64    &   64   &    64  &   70.3   &    0.60\\
HD133803   &  2016-06-27   &     80    &   64    &   80   &     64 & 125.8    &   1.11\\
$\beta$ Cir   &  2015-03-30   &   112    &   32    & 224   &    16  &   26.0    &   1.93\\
HD140840   &  2016-04-20   &     64    &   64    &   64    &    64  &   32.7   &    3.02\\
HD141378   &  2015-06-05   &     64    &   64    &   64    &    64  &   39.1   &    1.75\\
$\pi$ Ara      &  2016-06-10   &     80    &   64    &   80    &    64  &   24.9   &    0.50\\
HD174429   &  2014-07-15   &       2    &   60    &   6      &    20  &     7.6   &    0.88\\
HD178253   &  2015-06-06   &   128    &   32    & 256    &    16  &   48.8   &    1.44\\
$\eta$ Tel     &  2015-05-05   &     84    &   64    & 168    &    32  &   47.1   &   1.10\\
HD181327   &   2015-05-10   &    56    &   64    &    56   &    64  &   31.7    &   1.41\\
HD188228   &   2015-05-12   &   256   &   16    &  256   &    16  &   23.7    &   0.67\\
$\rho$ Aql    &   2015-09-27   &   128   &    32   &  256   &    16  &   25.0    &   1.50\\
HD202917   &   2015-05-31   &     64    &    64   &    64   &    64  &   49.5    &   1.35\\
HD206893   &   2016-09-15   &     80    &    64   &    80   &    64  &   76.2    &   0.63\\
HR8799       &   2014-07-13   &     20    &      8   &    40   &      4  &   18.1    &    0.8  \\
HD219482   &   2015-09-30   &    120   &   32   &   240   &   16   &   27.1   &   0.84\\
HD220825   &   2015-09-24   &    150   &   32   &   300   &   16   &   41.5   &   1.05\\
\hline  
\end{tabular}
\end{table*}

\subsection{Detection limits}

The contrast for each dataset was obtained using the procedure described in \cite{Zurlo1} and in \cite{Mesa}. The self-subtraction of the high contrast imaging methods adopted was evaluated by injecting simulated planets with known flux in the original datasets and reducing these data applying the same methods. \\
To translate the contrast detection limits into companion mass detection limits we used the theoretical model AMES-COND \citep{Baraffe} that is consistent with a hot-start planetary formation due to disk instability. These models predict lower planet mass estimates than cold-start models \citep{Marley} which, instead, represent the core accretion scenario, and affect considerably detection limits. We do not take into account the cold model since with such hypothesis we would not be able to convert measured contrasts in Jupiter masses close to the star. \cite{Spiegel} developed a compromise between the hottest disk instability and the coldest core accretion scenarios and called it warm-start. For young systems ($\le 100 $Myrs) the differences in mass and magnitude between the hot- and warm-start models are significant. Thus, the choice of the former planetary formation scenario has the implication of establishing a lower limit to detectable planet masses. We want to point out, however, that even if detection limits would be strongly influenced by the choice of warm- in place of hot-start models our dynamical conclusions (as obtained in the following sections) would not change significantly. Indeed, we would obtain in any case that for the great majority of the systems in our sample the gap between the belts and the absence of revealed planets would be explained only adding more than one planet and/or considering eccentric orbits.\\
In order to obtain detection limits in the form $a_{\mathrm{p}}$ vs. $M_{\mathrm{p}}$ we retrieved the J, H and Ks magnitudes from 2MASS and the distance to the system from Table \ref{tabu2}.
The age determination of the targets is based on the methodology described in \citet{Desidera} but adjusting the ages of several young moving groups to the latest results as in  \citet{Vigan}.\\
Eventually, the IFS and IRDIS detection limits were combined in the inner parts of the field of view (within 0.7 arcsec), adopting the lowest in terms of companions masses.
Detection limits obtained by such procedures are mono-dimensional since they depend only on the distance $a_P$ from the star. More precise bi-dimensional detection limits could be implemented to take into account the noise due to the luminosity of the disk and its inclination \citep[see for instance Figure 11 of][]{Rodet}. Whereas the disk's noise is for most systems negligible (it becomes relevant only for very luminous disks), the projection effects due to inclination of the disk could strongly influence the probability of detecting the planets. We will consider the inclination caveat when performing the analysis for two and three equal mass planets on circular orbits in Section 6.

\section{Dynamical predictions for a single planet}

\subsection{General Physics}
A planet sweeps an entire zone around its orbit that is proportional to its semi-major axis and to a certain power law of the ratio $\mu$ between its mass and the mass of the star.\\
One of the first to reach a fundamental result in this field was \cite{Wisdom}  who estimated the stability of dynamical systems for a non linear Hamiltonian with two degrees of freedom. Using the approximate criterion of the zero order resonance overlap for the planar circular-restricted three-body problem, he derived the following formula for the chaotic zone that surrounds the planet
\begin{equation}
\label{W}
\Delta a= 1.3 \mu^{2/7} a_p,
\end{equation} 
where $\Delta a$ is the half width of the chaotic zone, $\mu$ the ratio between the mass of the planet and the star, $a_p$ is the semi-major axis of the planet's orbit.\\
After this first analytical result, many numerical simulations have been performed in order to refine this formula. One particularly interesting expression regarding the clearing zone of a planet on a circular orbit was derived by \cite{Morrison}. The clearing zone, compared to the chaotic zone, is a tighter area around the orbit of the planet in which dust particles become unstable and from which are ejected rather quickly. The formulas for the clearing zones interior and exterior to the orbit of the planet are
\begin{equation}
\label{Mo1}
 (\Delta a)_{in}=1.2 \mu^{0.28} a_p
 \end{equation}
 \begin{equation}
 \label{Mo2}
 (\Delta a )_{ext}=1.7 \mu^{0.31} a_p.
 \end{equation}
The last result we want to highlight is the one obtained by \cite{Mustill2} using again N-body integrations and taking into account also the eccentricities $e$ of the particles. Indeed, particles in a debris disk can have many different eccentricities even if the majority of them follow a common stream with a certain value of $e$. The expression for the half width of the chaotic zone in this case is given by 
\begin{equation}
\label{Must}
\Delta a = 1.8 \mu^{1/5} e^{1/5} a_p.
\end{equation}
The chaotic zone is thus larger for greater eccentricities of particles. The equation \eqref{Must} is only valid for values of $e$ greater than a critical eccentricity, $e_{crit}$, given by
\begin{equation}
 e_{crit}\sim0.21 \mu^{3/7}.
 \end{equation}
 For $e<e_{crit}$ this result is not valid anymore and equation \eqref{W} is more reliable. Even if each particle can have an eccentricity due to interactions with other bodies in the disk, such as for example collisional scattering or disruption of planetesimals in smaller objects resulting in high $e$ values, one of the main effects that lies beneath global eccentricity in a debris disk is the presence of a planet on eccentric orbit. \cite{Mustill} have shown that the linear secular theory gives a good approximation of the forced eccentricity $e_f$ even for eccentric planet and the equations giving $e_f$ are:  
 \begin{equation}
 \label{ei}
 e_{f,in}\sim\frac{5ae_p}{4a_p}
 \end{equation}
 \begin{equation}
 \label{ee}
 e_{f,ex}\sim\frac{5a_pe_p}{4a},
 \end{equation} 
 where $e_{f,in}$ and $e_{f,ex}$ are the forced eccentricities for planetesimals populating the disk interior and exterior to the orbit of the planet, respectively; $a_p$ and $e_p$ are, as usual, semi-major axis and eccentricity of the planet, while $a$ is the semi-major of the disk. We note that such equations arise from the leading-order (in semi-major axis ratio) expansion of the Laplace coefficients, and hence are not accurate when the disk is very close to the planet.\\
It is of common use to take the eccentricities of the planet and disk as equal, because this latter is actually caused by the presence of the perturbing object. Such approximation is also confirmed by equations \eqref{ei} and \eqref{ee}. Indeed, the term $5/4$ is balanced by the ratio between the semi-major axis of the planet and that of the disk since, in our assumption, the planet gets very close to the edge of the belt and thus the values of $a_p$ are not so different from that of $a$, giving $e_{f}\sim e_p$. \\
 Other studies, such as the ones presented in \cite{Chiang} and \cite{Quillen}, investigate the chaotic zone around the orbit of an eccentric planet and they all show very similar results to the ones discussed here above. However, in no case the eccentricity of the planet appears directly into analytical expressions, with the exception of the equations presented in \cite{Pearce} that are compared with our results in Section 5.2.

\subsection{Numerical simulations}
\begin{figure} [h!]
\includegraphics[scale=1.3]{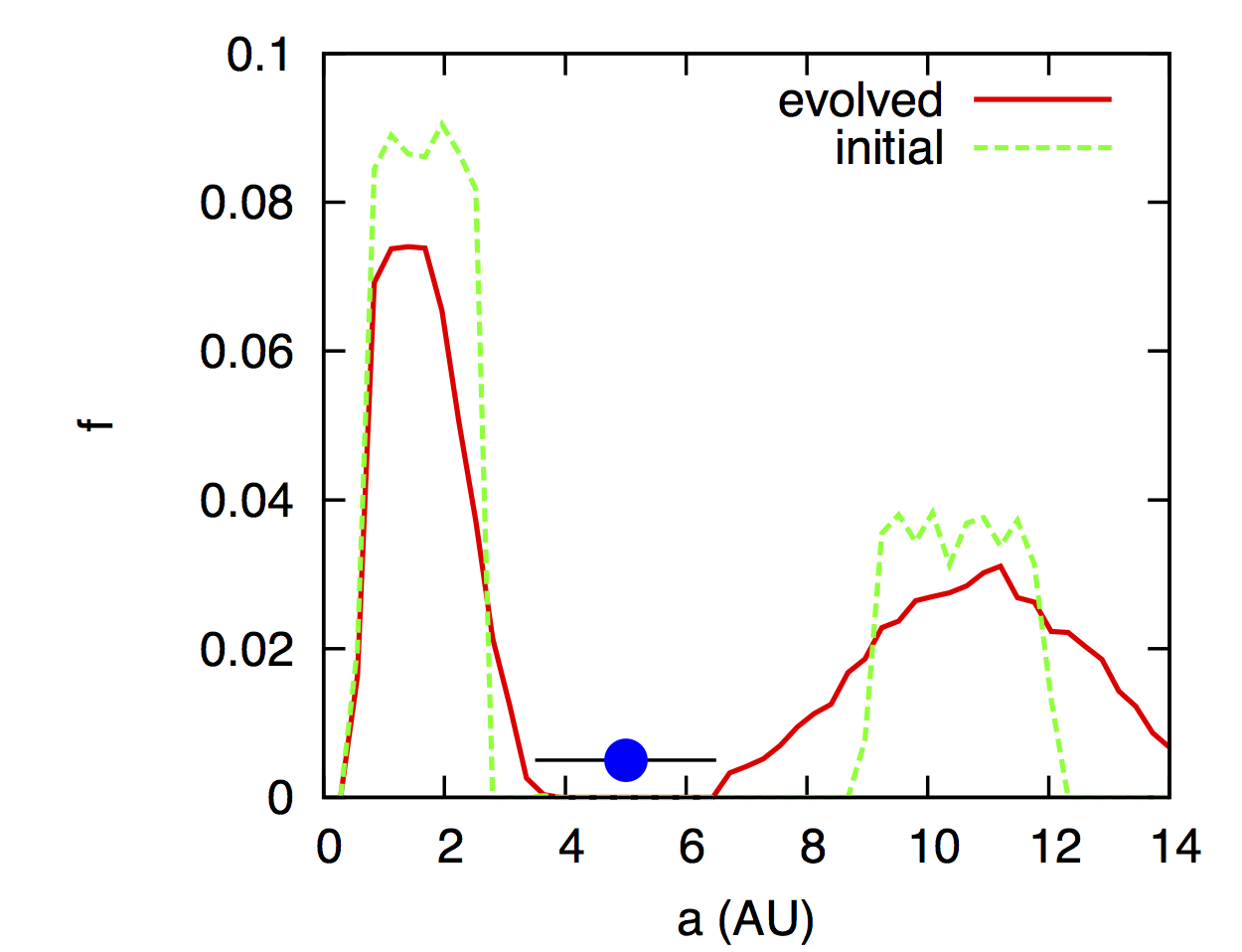}
\caption{Numerical simulation for a planet of $1$ $M_J$ around a star of $1$ $M_{\odot}$ with a semi-major axis of $5$ AU and eccentricity of $0.3$. We plot the fraction of bodies that are not ejected from the system as a function of the radius. Green lines represent the stability analysis on the radial distribution of the disk. Red lines represent the radial distributions of $4000$ objects.}
\label{Simu}
\end{figure}

\begin{figure} 
\begin{tabular} {@{}c@{}}
\includegraphics[width=0.45\textwidth]{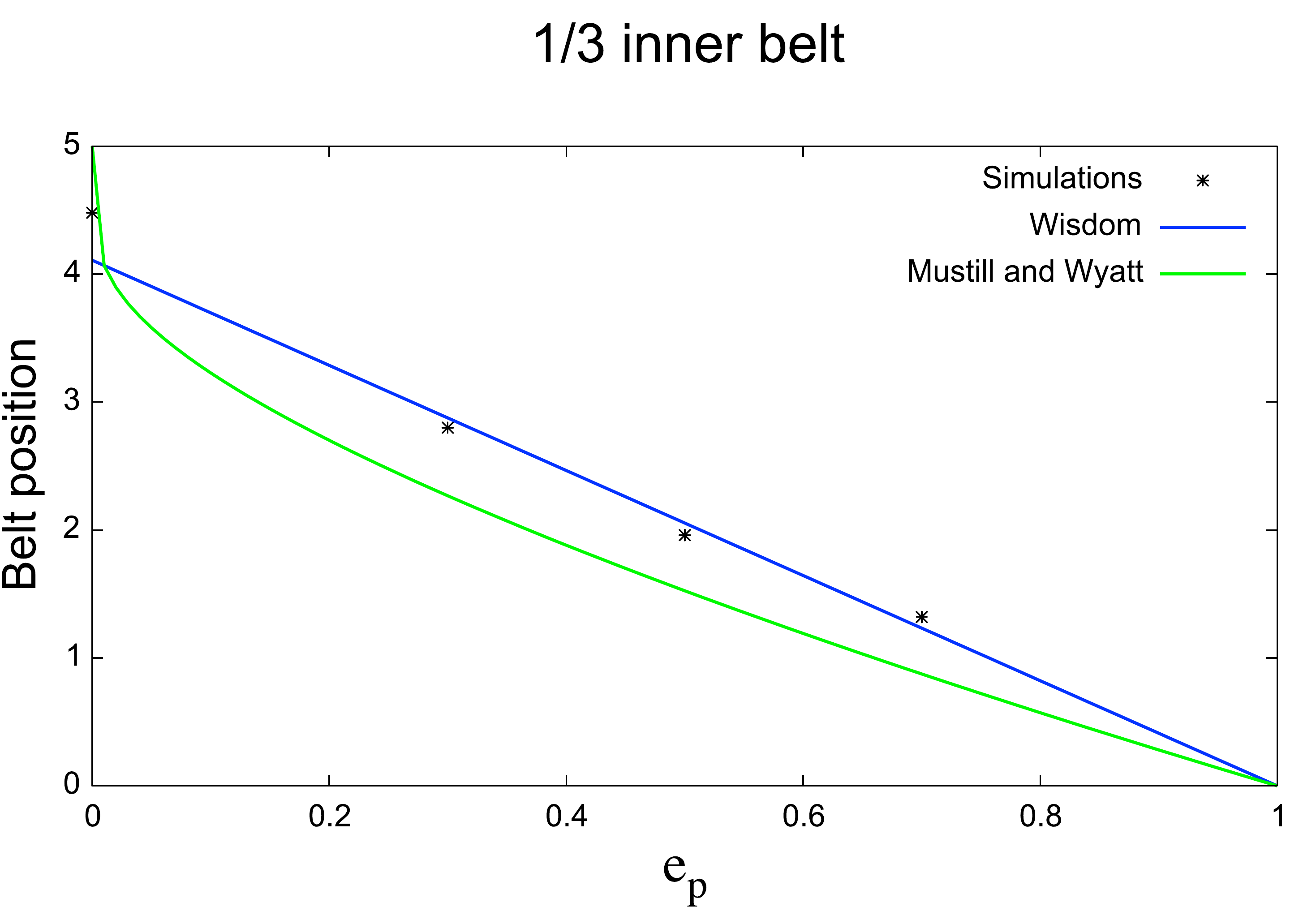}\\
\end{tabular}
\begin{tabular} {@{}c@{}}
\includegraphics[width=0.45\textwidth]{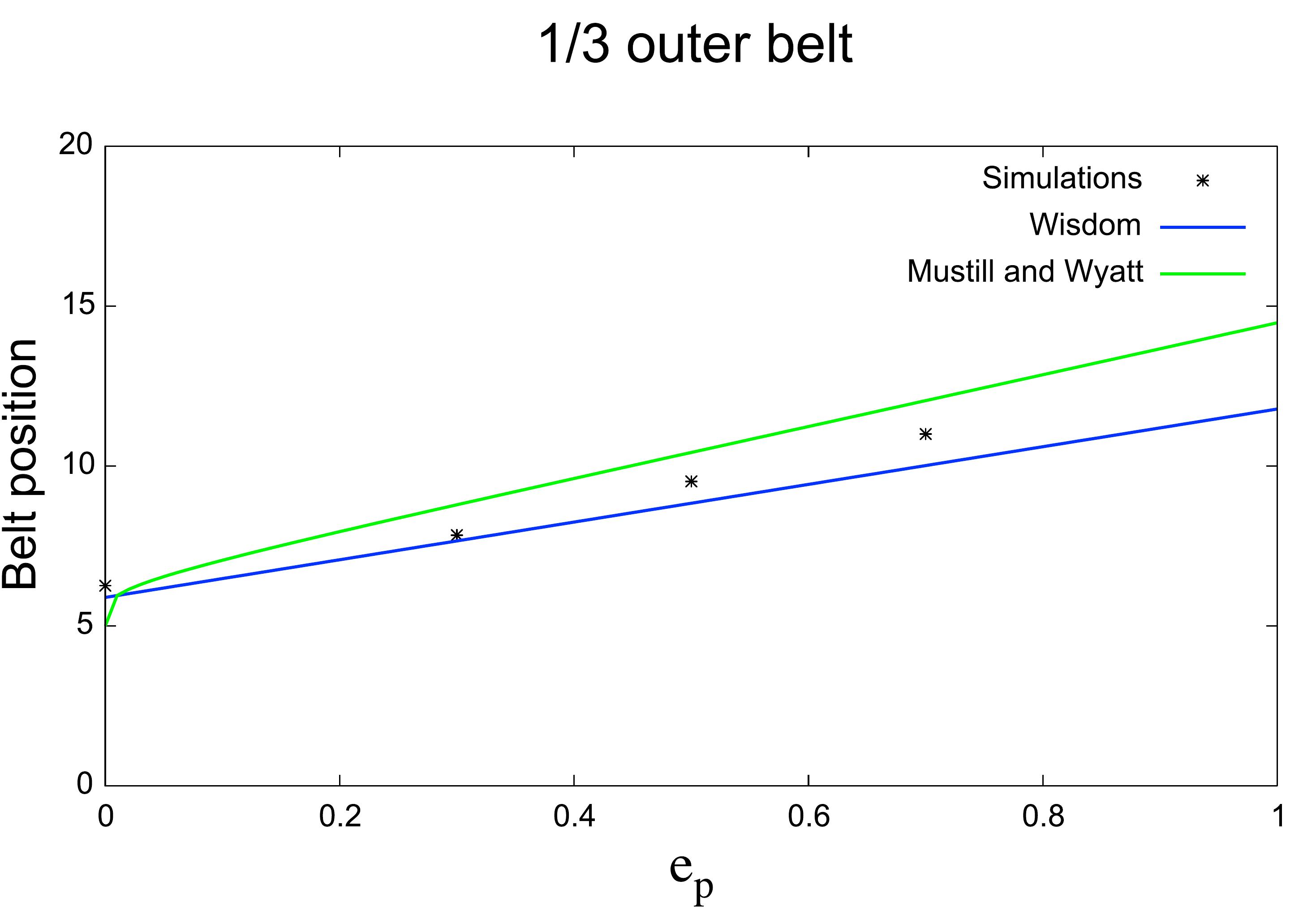}\\
\end{tabular}
\caption{Position of the inner (up) and external (down) belt for cuts off of  $1/3$.}
\label{S}
\end{figure}

All the previous equations apply to the chaotic zone of a planet moving on a circular orbit. When we introduce an eccentricity $e_p$ the planet varies its distance from the star. Recalling all the formulations for the clearing zone presented in Section 5.1 we can see that it always depends on the mean value of the distance between the star and the planet, $a_p$, that is assumed to be almost constant along the orbit.\\
The first part of our analysis considers one single planet as the only responsible for the lack of particles between the edges of the inner and external belts. We will show that the hypothesis of circular motion is not suitable for any system when considering masses under $50$ $M_J$.
For this reason, the introduction of eccentric orbits is of extreme importance in order to derive a complete scheme for the case of a single planet.\\
In order to account for the eccentric case we have to introduce new equations. The empirical approximation that we will use (and that we will support with numerical simulations) consists in starting from the old formula \eqref{W} and \eqref{Must} suitable for the circular case and replace the mean orbital radius, $a_p$, with the positions of apoastron, $Q$, and periastron, $q$, in turn. We thus get the following equations
\begin{equation}
\label{apow}
(\Delta a)_{ex}= 1.3 \mu^{2/7} Q
\end{equation}
\begin{equation}
\label{periw}
(\Delta a)_{in}= 1.3 \mu^{2/7} q,
\end{equation}
which replace Wisdom formula \eqref{W}, and
\begin{equation}
\label{apom}
(\Delta a)_{ex} = 1.8 \mu^{1/5} e^{1/5} Q
\end{equation}
\begin{equation}
\label{perim}
(\Delta a )_{in}= 1.8 \mu^{1/5} e^{1/5} q,
\end{equation}
which replace Mustill \& Wyatt's equation \eqref{Must} and in which we choose to take as equal the eccentricities of the particles in the belt and that of the planet, $e=e_p$.
The substitution of $a_p$ with apoastron and periastron is somehow similar to consider the planet as split into two objects, one of which is moving on circular orbit at the periastron and the other on circular orbit at apoastron and both with mass $M_p$.\\
In parallel with this approach, we performed a complementary numerical investigation of the location of the inner and outer limits of the gap carved in a potential planetesimal disk by a massive and eccentric planet orbiting within it to test the reliability of the analytical estimations in the case of eccentric planets. \\
We consider as a test bench a typical configuration used in numerical simulations with a planet of $1$ $M_J$ around a star of $1$ $M_{\odot}$ and two belts, 
one external and one internal to the orbit of the planet, composed of massless objects. The planet has a semi-major axis of $5$ AU and eccentricities of $0$, $0.3$, $0.5$ and $0.7$ in the four simulations. We first perform a stability 
analysis of random sampled orbits using the frequency map analysis (FMA) \citep{Marzari}. A large number of putative planetesimals are generated with semi-major axis $a$ uniformly distributed within the intervals $a_p+ R_H < a < a_p + 30 R_H$ and $a_p- 12 R_H < a < a_p - R_H$, where $R_H$ is the radius of the Hill's sphere of the planet. The initial eccentricities are small (lower than $0.01$) so that the planetesimals acquire a proper eccentricity equal to the one forced by the secular perturbations of the planet. This implicitly assumes that the planetesimal belt was initially in a cold state.\\ 
The FMA analysis is performed on the non-singular variables $h$ and $k$, defined as $h = e \sin{\varpi}$ and $k = e \cos{\varpi}$, of each planetesimal in the sample. The main frequencies present in the signal are due to the secular
perturbations of the planet. Each dynamical system composed of planetesimal, planet and central star is numerically
integrated for $5$ Myr with the RADAU integrator and the FMA analysis is performed using running time windows
extending for $2$ Myr. The main frequency is computed with the FMFT high precision algorithm described in \cite{Laskar} and \cite{Sidlichovsky}. The chaotic diffusion of the orbit is measured as the logarithm of the relative change of the main frequency of the signal over all the windows, $c_s$. The steep decrease in the value of $c_s$ marks the onset of long term stability for the planetesimals and it outlines the borders of the gap sculpted by the planet. \\
This approach allows a refined  determination of the half width of the chaotic zone for eccentric planets. We term the inner and outer values of semi-major axis of the gap carved by the planet in the planetesimal disk $a_i$ and $a_o$, respectively. For $e=0$, we retrieve the values of $a_i$ and $a_o$ that can be derived from eq. \eqref{W} even if $a_o$ is slightly larger in our model ($6.1$ AU instead of $5.9$ AU). For increasing values of the planet eccentricity, $a_i$ moves inside while $a_o$ shifts outwards, both almost linearly. However, this trend is in semi-major axis while the spatial distribution of planetesimals depends on their radial distance. \\
For increasing values of $e_p$, the planetesimal eccentricities grow as predicted by equations \eqref{ei} and \eqref{ee} and the periastron of the planetesimals in the exterior disk moves inside while the apoastron in the interior disk moves outwards. As a consequence, the radial distribution trespasses $a_i$ and $a_o$ reducing the size of the gap. To account for this effect, we integrated the orbits of $4000$ planetesimals for the interior disk and just as many for the exterior disk. \\
The bodies belonging to the exterior disk are generated with semi-major axis $a$ larger than $a_o$ while for the interior disk $a$ is smaller than $a_i$. After a period of $10$ Myr, long enough for their pericenter longitudes to be randomized, we compute the radial distribution. This will be determined by the eccentricity and periastron distributions of the planetesimals forced by the secular perturbations. In Figure \ref{Simu} we show the normalized radial distribution for $e_p = 0.3$. At the end of the numerical simulation the radial distribution extends inside $a_o$ and outside $a_i$. \\ 
The inner and outer belts are detected by the dust produced in collisions between the planetesimals. There are additional forces that act on the dust, like the Poynting-Robertson drag, slightly shifting the location of the debris disk compared to the radial distribution of the planetesimals. However, as a first approximation, we assume that the associated dusty disk coincides with the location of the planetesimals. In this case the outer and inner borders $d_2$ and $d_1$ of the external and internal disk, respectively, can be estimated as the values of the radial distance for which the density distribution of planetesimals drop to $0$. In alternative, we can require that the borders of the disk are defined where the dust is bright enough to be detected and this may occur when the radial distribution of the planetesimals is larger than a given ratio of the peak value, $f_M$, in the density distribution.\\
We arbitrarily test two different limits, one of $1/3$  $f_M$ and the other of $1/4$ $f_M$, for both the internal and external disks. In this way, the low density wings close to the planet on both sides are cut away under the assumption that they do not produce enough dust to be detected. \\
The $d_2$ and $d_1$ outer and inner limits of the external and internal disks are given in all three cases ($0$, $1/3$ and $1/4$) in Table \ref{tab2} and \ref{tab3} for each eccentricity tested. The first three columns report the results of our simulations and are compared to the estimated values of the positions of the belts (last two columns) that we obtained in first place, calculating the half width of the chaotic zone from equations \eqref{periw} and \eqref{perim} for the inner belt and \eqref{apow} and \eqref{apom} for the outer one, and then we use the relations
\begin{equation}
(\Delta a)_{in}= q - d_1
\label{ain}
\end{equation}
\begin{equation}
(\Delta a)_{ex}=d_2 -Q
\label{aex}
\end{equation}
obtaining, in the end, $d_1$ and $d_2$. \\
We plot the positions of the two belts against the eccentricity for cut off of $1/3$ in Figure \ref{S}. As we can see, results from  simulations are in good agreement with our approximation. Particularly, we note how Wisdom is more suitable for eccentricities up to $0.3$ (result that has already been proposed in a paper by \cite{Quillen} in which the main conclusion was that particles in the belt do not feel any difference if there is a planet on circular or eccentric orbit for $e_p\le 0.3$). For greater values of $e_p$ also equations \eqref{apom} and \eqref{perim} give reliable results. \\
In \cite{Pearce} a similar analysis is discussed for what regards the inner edge of a debris disk due to an eccentric planet that orbits inside the latter. As known from the second Kepler's law, the planet has a lower velocity when it orbits near apoastron and thus it spends more time in such regions. Thus, they assumed that the position of the inner edge is mainly influenced by scattering of particles at apocenter in agreement with our hypothesis. Using the Hill stability criterion they obtain for the chaotic zone the following expression
\begin{equation}
\Delta a_{ex}= 5 R_{H, Q},
\label{pearce}
\end{equation}
where $R_{H,Q}$ is the Hill radius for the planet at apocenter, given by
\begin{equation}
R_{H, Q}\sim a_p(1+e_p)\Bigl[\frac{M_p}{(3-e_p)M_*}\Bigr]^{1/3}.
\label{hill}
\end{equation}
Comparing $\Delta a_{ext}$ to equations \eqref{apow}, \eqref{apom} and \eqref{pearce} for a planet of 1 $M_J$ that orbits around a star of 1 $M_{\odot}$ with a semi-major axis of $a_p=5 AU$ (that are the typical values adopted in \cite{Pearce}) and eccentricity $e_p=0.3$, we obtain results that are in good agreement and differ by $15\%$. For the same parameters but higher eccentricity ($e_p=0.5$) the difference steeply decreases down to $2\%$.  Thus, even if our analysis is based on different equations with respect to \eqref{pearce} the clearing zone that we obtain is in good agreement with values as expected by \cite{Pearce}, giving a farther corroboration of our assumption for planets on eccentric orbits.\\

\begin{table}
\caption{Position of the inner belt}
\label{tab2}
\centering
\begin{tabular}{ccccc}
\hline\hline
$e_p$ & cut& \bf{$d_{1,num}$(AU)}  & Wisdom & Mustill\\
\hline  
 0&0&4.1&4.11&4.11\\
0& $1/3$ & 4.48 & 4.11 & 4.11\\
0& $1/4$ & 4.48 & 4.11 & 4.11\\
0.3&0&2.5& 2.88&2.27\\
0.3&$1/3$&2.8& 2.88&2.27\\
0.3&$1/4$&3.1&2.88&2.27\\
0.5&0&1.74&2.05&1.53\\
0.5&$1/3$&1.96&2.05&1.53\\
0.5&$1/4$&2.24&2.05&1.53\\
0.7&0&1.1&1.23&0.87\\
0.7&$1/3$&1.32&1.23&0.87\\
0.7&$1/4$&1.38&1.23&0.87\\
\hline   
\end{tabular}
\end{table}

\begin{table}
\caption{Position of the outer belt}
\label{tab3}
\centering
\begin{tabular}{ccccc}
\hline\hline
$e_p$ & cut& \bf{$d_{2,num}$(AU)}  & Wisdom & Mustill\\
\hline  
0&0&6.1& 5.89 & 5.89\\
0& $1/3$ & 6.26 & 5.89 & 5.89\\
0& $1/4$ & 6.26 & 5.89 & 5.89\\
0.3&0&8.9& 7.66&8.79\\
0.3&$1/3$&7.84& 7.66&8.79\\
0.3&$1/4$&7.56&7.66&8.79\\
0.5&0&10.27&8.84&10.42\\
0.5&$1/3$&9.52&8.84&10.42\\
0.5&$1/4$&9.24&8.84&10.42\\
0.7&0&11.6&10.02&12.05\\
0.7&$1/3$&11.0&10.02&12.05\\
0.7&$1/4$&10.79&10.02&12.05\\
\hline  
\end{tabular}
\end{table}

\subsection{Data analysis}
\begin{figure} [t] 
\centering
\begin{tabular}{@{}c@{}}
\includegraphics[width=0.45\textwidth]{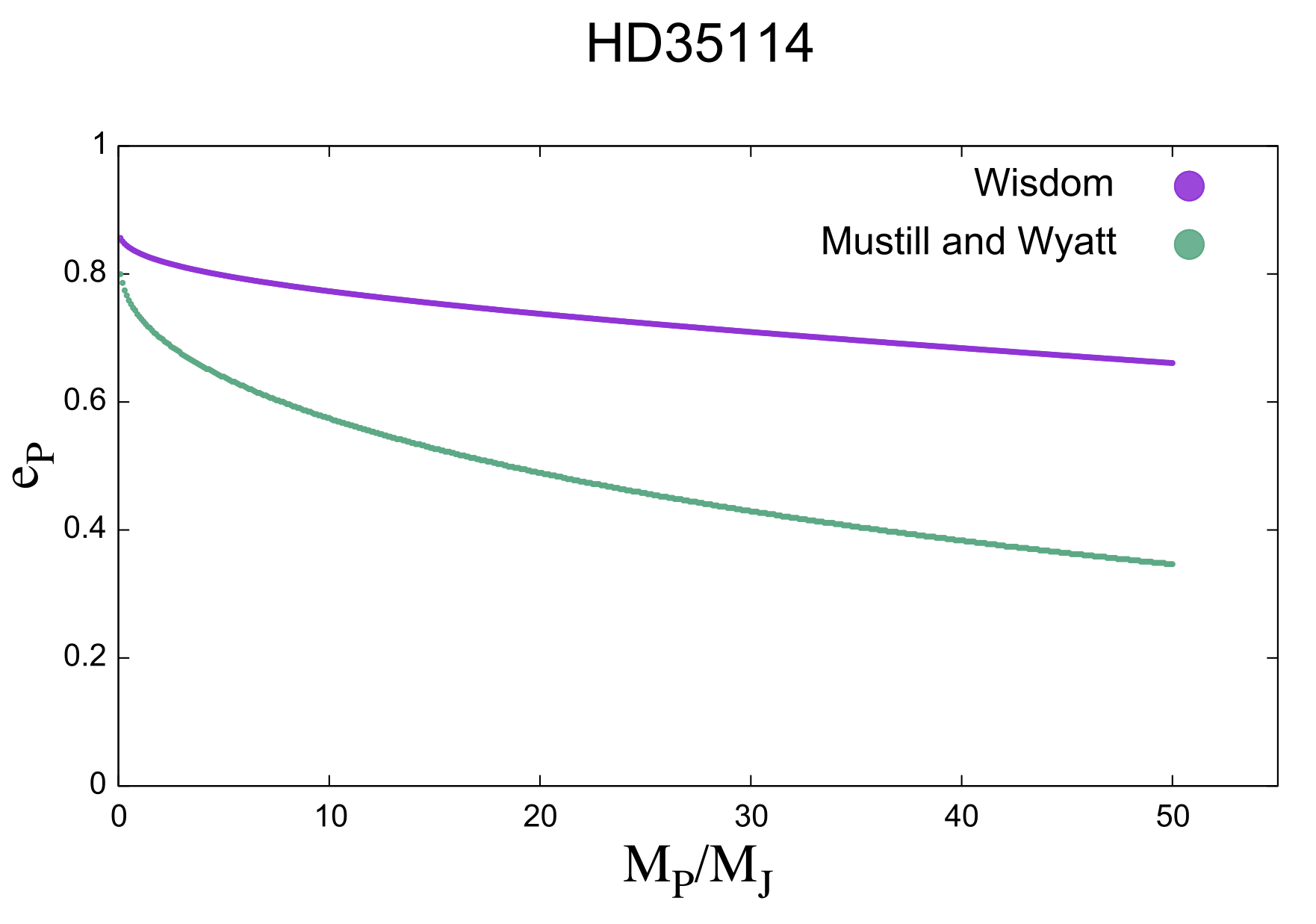}\\
\end{tabular}
\begin{tabular}{@{}c@{}}
\includegraphics[width=0.45\textwidth]{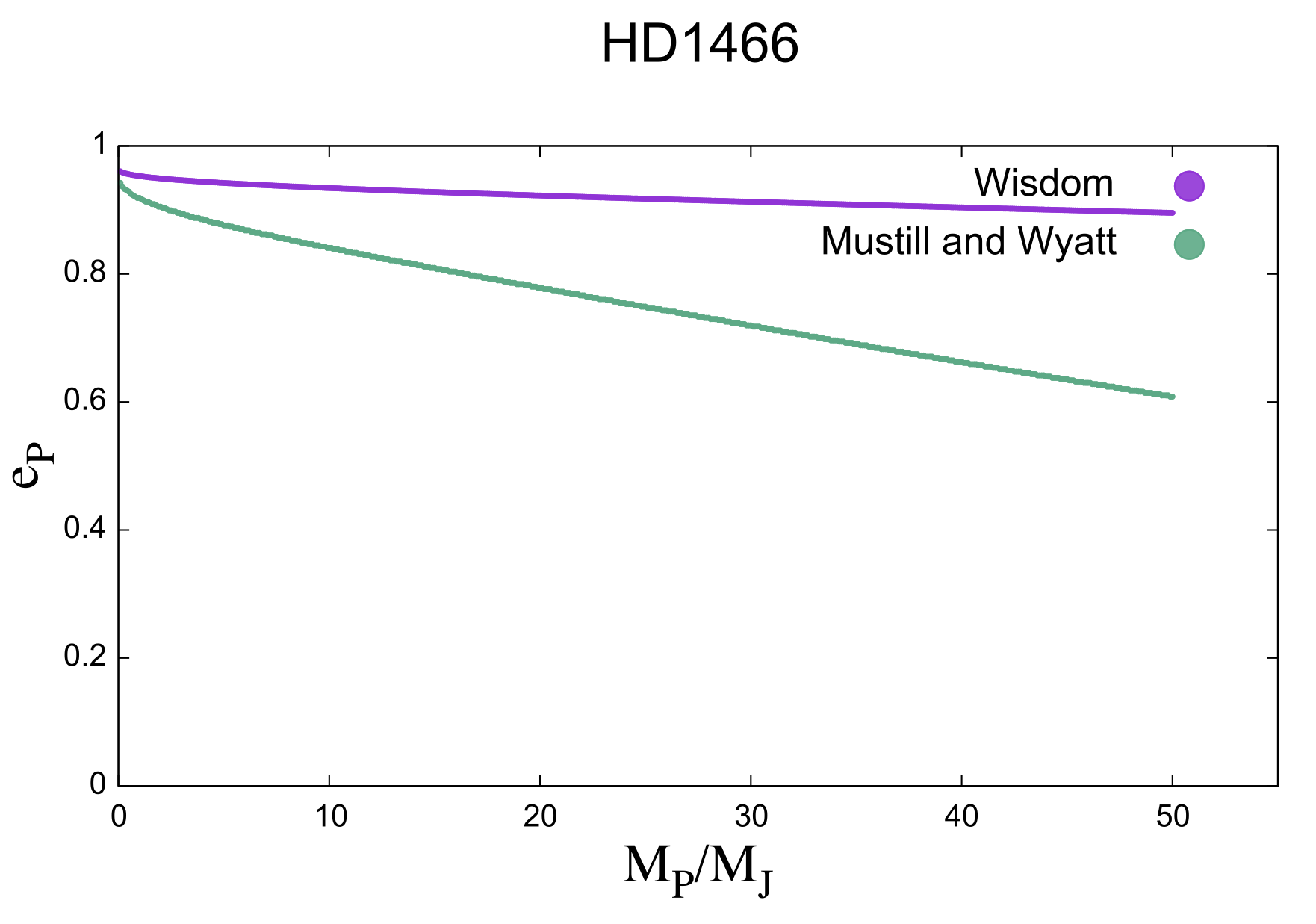}\\
\end{tabular}
\caption{ $e_p$ versus $M_p$ for HD35114 (up) and HD1466 (down) }
\label{eM}
\end{figure}

\begin{figure} [h!] 
\centering
\begin{tabular}{@{}c@{}} 
\includegraphics[width=0.45\textwidth]{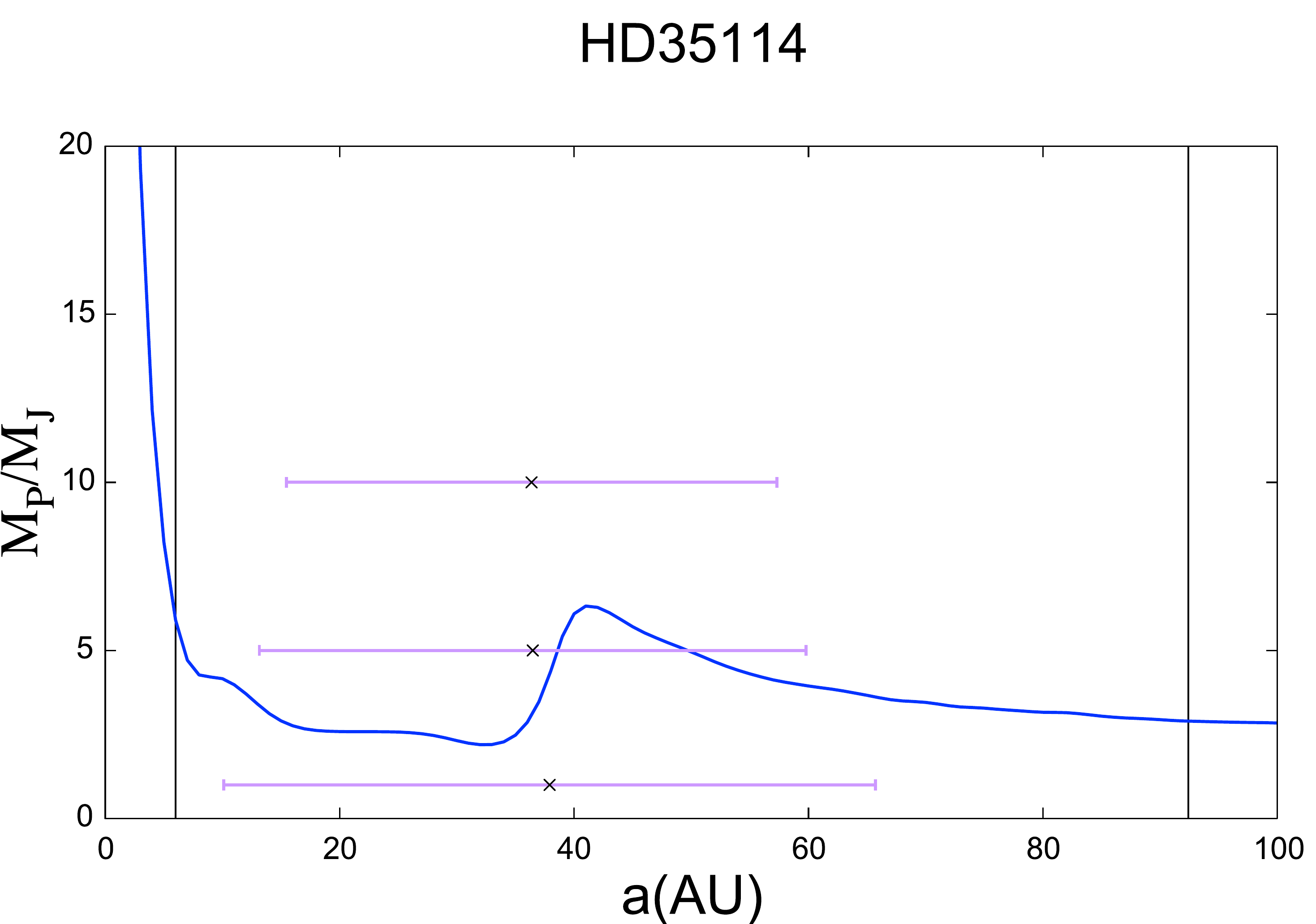}\\
\end{tabular}
\begin{tabular}{@{}c@{}} 
\includegraphics[width=0.45\textwidth]{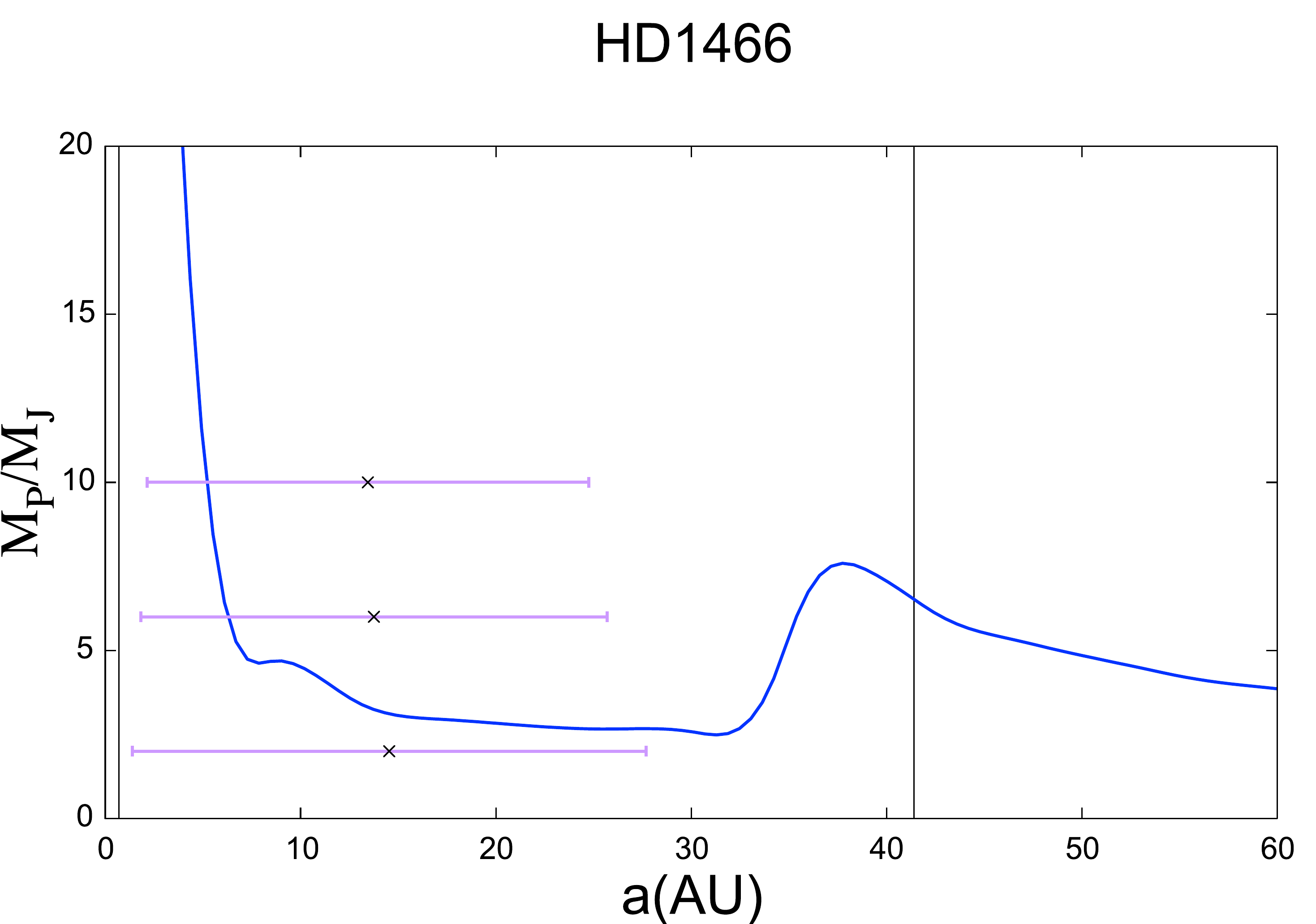}\\
\end{tabular}
\caption{SPHERE detection limits for HD35114 (up) and HD1466 (down). The bar plotted for each $M_p$ represents the interval of distances covered by the planet during its orbit, from a minimum distance (periastron) to the maximum one (apoastron) from the star. The two vertical black lines represent the positions of the two belts. Projection effects in case of significantly inclined systems are not included.}
\label{HD}
\end{figure}

Once we have verified the reliability of our approximations, we proceed analyzing the dynamics of the systems in the sample.\\
The first assumption that we tested is of a single planet on a circular orbit around its star. We use the equations for the clearing zone of Morrison \& Malhotra \eqref{Mo1} and \eqref{Mo2}. We vary the mass of the planet between $0.1$ $M_J$, i. e. Neptune/Uranus sizes, and $50$ $M_J$ in order to find the value of $M_p$, and the corresponding value of $a_p$, at which the planet would sweep an area as wide as the gap between the two belts. $50$ $M_J$ represents the approximate upper limit of applicability of the equations, since they require that $\mu$ is much smaller than $1$.  Since $(\Delta a)_{in}+(\Delta a)_{ex}= d_2 -d_1$, knowing $M_p$, we can obtain the semi-major axis of the planet by
\begin{equation}
a_p=\frac{d_2-d_1}{1.2\mu^{0.28}+1.7\mu^{0.31}}.
\end{equation}
With these starting hypothesis we cannot find any suitable solution for any system in our sample. Thus objects more massive than $50$ $M_J$ are needed to carve out such gaps but they would  clearly lie well above the detection limits.\\
Since we get no satisfactory results for the circular case, we then consider one planet on eccentric orbit. We use the approximation illustrated in the previous paragraph with one further assumption: we consider the apoastron of the planet as the point of the orbit nearest to the external belt while the periastron as the nearest point to the inner one. We let again the masses vary in the range $[0.1,50]$ $M_J$ and, from equations \eqref{apow}, \eqref{periw}, \eqref{apom} and \eqref{perim}, we get the values of periastron and apoastron for both Wisdom and Mustill \& Wyatt formulations, recalling also equations \eqref{ain} and \eqref{aex}. Therefore, we can deduce the eccentricity of the planet through 
\begin{equation}
e_p=\frac{Q-q}{Q+q}.
\end{equation}
The equation \eqref{Must} contains itself the eccentricity of the planet $e_p$, that is our unknown. The expression to solve in this case is 
\begin{equation}
e_p-\frac{d_2(1-1.8(\mu e_p)^{1/5})-d_1(1+1.8(\mu e_p)^{1/5})}{d_2(1-1.8(\mu e_p)^{1/5})+d_1(1+1.8(\mu e_p)^{1/5})}=0
\label{zero}
\end{equation}
for which we found no analytic solution but only a numerical one.
We can now plot the variation of the eccentricity as a function of the mass. We present two of these graphics, as examples, in Figure \ref{eM}.\\
In each graphic there are two curves one of which represents the analysis carried out with the Wisdom formulation and the other with Mustill \& Wyatt expressions. In both cases, the eccentricity decreases with increasing planet mass. This is an expected result since a less massive planet has a tighter chaotic zone and needs to come closer to the belts in order to separate them of an amount $d_2-d_1$, that is fixed by the observations (and viceversa for a more massive planet that would have a wider $\Delta a$). Moreover, we note that the curve that represents Mustill \& Wyatt's formulas decreases  more rapidly than Wisdom's curve does. This is due to the fact that equation \eqref{Must} takes also into account the eccentricity of the planetesimals (in our case $e=e_p$) and thus $\Delta a$ is wider.\\
Comparing the graphics of the two systems, representative of the general behavior of our targets, we note that whereas for HD35114, for increasing mass, the eccentricity reaches intermediate values ($\le 0.4$), HD1466 needs planets on very high eccentric orbits even at large masses ($\ge0.6$). From Table \ref{tabu2}, the separation between the belts in HD35114 is of $\sim86$ AU whereas in HD1466 is only $\sim40$ AU. We can then wonder why in the first system planets with smaller eccentricity are needed to dig a gap larger than the one in the second system. The explanation regards the positions of the two belts: HD35114 has the inner ring placed at $6$ AU whereas HD1466 at $0.7$ AU. From equations \eqref{W} and \eqref{Must} we obtain a chaotic zone that is larger for further planets since it is proportional to $a_p$.\\
From the previous discussion, we deduce that many factors in debris disks are important in order to characterize the properties of the planetary architecture of a system, first of all the radial extent of the gap between the belts, the wider the more massive and/or eccentric planets needed, but also the positions of the belts (the closer to the star, the more difficult to sculpt) and the mass of the star itself.\\
 For most of our systems the characteristics of the debris disks are not so favorable to host one single planet since we would need very massive objects that have not been detected. For this reason we now analyze the presence of two or three planets around each star.\\
 Before considering multiple planetary systems, however, we want to compare our results with the detection limits available in the sample and obtained as described in Section 4.2.  We show, as an example, the results for HD35114 and HD1466 in Figure \ref{HD} in which we plot the detection limits curve, the positions of the two belts (the vertical black lines) and three values of the mass. From the previous method we can associate to each value of the mass a value of $a_p$ and $e_p$ noting that, as mentioned above, Wisdom gives more reliable results for $e_p\le0.3$ whereas Mustill \& Wyatt for $e_p>0.3$.    
Moreover, we choose three values of masses because they represent well the three kinds of situations that we could find: for the smallest mass the planet is always below the detection limits and so never detectable; for intermediate mass the planet crosses the curve and thus it is at certain radii of its orbit detectable and at others undetectable (we note however that the planet spends more time at apoastron than at periastron so that it is more likely detectable in this latter case); the higher value of $M_p$ has only a small portion of its orbit (the area near to the periastron) that is hidden under the curve and thus undetectable.\\
We note that in both figures there is a bump in the detection limits curves: this is due to the passage from the deeper observations done with IFS that has field of view ($\le 0.8$ arcsec) to the IRDIS ones that are less deep but cover a greater range of distances (up to $5.5$ arcsec).

\section{Dynamical predictions for two and three planets}
\subsection{General physics}
In order to study the stability of a system with two planets, we have to characterize the region between the two. From a dynamical point of view, this area is well characterized by the Hill criterion.
Let us consider a system with a star of mass $M_*$, the inner planet with mass $M_{p,1}$, semi-major axis of $a_{p,1}$ and eccentricity $e_{p,1}$, and the outer one with mass $M_{p,2}$, semi-major axis $a_{p,2}$ and eccentricity $e_{p,2}$. In the hypothesis of small planet masses, i.e. $M_{p,1}<<M_*$, $M_{p,2}<<M_*$ and $M_{p,2}+M_{p,1}<<M_*$, the system will be Hill stable \citep{Gladman} if
\begin{equation}
\label{stab}
\alpha^{-3}\Bigl(\mu_1+\frac{\mu_2}{\delta^2}\Bigr)(\mu_1\gamma_1+\mu_2\gamma_2\delta)^2\ge1+3^{4/3}\frac{\mu_1\mu_2}{\alpha^{4/3}},
\end{equation}
where $\mu_1$ and $\mu_2$ are the ratio between the mass of the inner/outer planet and the star respectively, $\alpha=\mu_1+\mu_2$, $\delta=\sqrt{1+\Delta/a_{p,1}}$ with $\Delta=a_{p,2}-a_{p,1}$ and, at the end, $\gamma_i=\sqrt{1-e_{p,i}^2}$ with $i=1,2$.\\
If the two planets in the system have equal masses, the previous equation, taking  $M_{p,2}=M_{p,1}=M_p$ and $\mu=M_p/M_*$, can be rewritten in the form
\begin{equation}
\label{2e}
\alpha^{-3}\Bigl(\mu+\frac{\mu}{2\delta^2}\Bigr)(\mu\gamma_1+\mu\gamma_2\delta)^2-1-3^{4/3}\frac{\mu^2}{\alpha^{4/3}}\ge 0
\end{equation}
and substituting the expressions for $\alpha$, $\delta$, $\Delta$ and $\gamma_i$ we obtain
\begin{equation}
\label{mass}
\frac{1}{8}\biggl(1+\frac{a_{p,1}}{a_{p,2}}\biggr)\biggl(\sqrt{1-e_{p,1}^2}+\sqrt{1-e_{p,2}^2}\sqrt{\frac{a_{p,2}}{a_{p,1}}}\biggr)^2-1-\biggl(\frac{3}{2}\biggr)^{4/3}\mu^{2/3}\ge0.
\end{equation}
Thus, the dependence of the stability on the mass of the two planets, in the case of equal masses, is very small since it appears only in the third term of the previous equation in the form $\mu^{2/3}$, with $ \mu << 1$ and $\mu \ge 0$, and, for typical values, it is two orders of magnitude smaller than the first two terms. The leading terms that determine the dynamics of the system are the eccentricities $e_{p,1}$ and $e_{p,2}$. For this reason, we expect that small variation in the eccentricities will lead to great variation in masses.\\
A further simplification to the problem comes when we consider two equal-mass planets on circular orbits. In this case the stability equation \eqref{stab} takes the contracted form 
\begin{equation}
\label{circ}
\Delta \ge 2\sqrt{3} R_H,
\end{equation}
where $\Delta$ is the difference between the radii of the planets' orbits and $R_H$ is the planets mutual Hill radius that, in the general situation, is given by
\begin{equation}
\label{hill}
R_H=\Bigl(\frac{M_{p,1}+M_{p,2}}{3M_*} \Bigr)^{1/3}\Bigl(\frac{a_{p,1}+a_{p,2}}{2}\Bigr).
\end{equation}
In the following, we will investigate both the circular and the eccentric cases with two planets of equal masses.\\
The last case that we present is a system with three coplanar and equal-mass giant planets on circular orbits. The physics follows from the previous discussion since the stability zone between the first and the second planet, and between the second and the third is again well described by the Hill criterion. Once fixed the inner planet semi-major axis $a_{p,1}$, the semi-major axis of the second and third planets are given by
\begin{equation}
\label{semi}
a_{p,i+1}=a_{p,i}+KR_{H i,i+1}
\end{equation}
where the $K$ value that ensure stability is a constant that depends on the mass of the planets and $R_{H i,i+1}$ is the mutual Hill radius between the first and the second planets for $i=1$ and between the second and the third for $i=2$. $K$ produces parametrizations curves, called K-curves, that are weakly constrained. However, we can associate to $K$ likely values that give us a clue on the architecture of the system. Following \cite{Marzari}, the most used values of $K$ for giant planets are:
\begin{itemize}
\item $K\sim 8$ for Neptune-size planets;
\item $K\sim 7$ for Saturn-size planets;
\item  $K\sim 6$ for Jupiter-size planets .
\end{itemize}
There is no analysis in literature that gives analytical tools to explore the case of three or more giant planets with different masses and/or eccentric orbits. Thus, it would be worth doing further investigations even if they go beyond the scope of this work.\\
As we will see in the following sections, once we have established the stability of a multi-planetary system we apply again the equations for the chaotic/clearing zone derived previously for a single planet as a criterion to describe the planet-disk interaction. However, for two and three planets more complex dynamical effects 
due to mean motion and secular resonances may change the expected positions of the edges. 
We have compared our analytical predictions with the results obtained by \cite{Moro1} who performed 
 numerical simulations in four systems (HD128311, HD202206, HD82943 and HR8799) with known companions in order to determine the positions of the gap. While the outer edge of the inner belt is well reproduced by the formulas we have exploited, the inner edge of the outer belt is slightly shifted farther out for each system in the numerical modeling. 
This is in part related to the stronger and more stable mean motion resonances in the single planets case. A full 
investigation of this problem is complex since the parameter space is wide as both the  mass and eccentricity of 
the planets may change. 
However, we are interested in a first order study and the differences due to the dynamical models are compatible with the error bars on the positions of the belts. Since we want only to give a method to obtain a rough estimation of possible architectures of planetary systems, such corrections will not be included in this paper but we stress that deeper analysis are needed to obtain stronger and more precise conclusions. 

\subsection{Data analysis}
\subsubsection{Two and three planets on circular orbits}
The first kind of analysis that we perform consists in taking into account two coplanar planets on circular orbits. In this case, between the two belts the system is divided into three different zones from  a stability point of view. The first one extends from the outer edge of the internal disk to the inner planet and it is determined from interaction laws between two massive bodies (the star and the planet) and N massless objects. The second zone is included between the inner and the outer planets and is dominated by the Hill's stability. Eventually, the third zone goes from the outer planet to the inner edge of the external belt and is an analog of the first one.\\
From equation \eqref{circ} we note that a system with two planets is stable if $\Delta=a_{p,2}-a_{p,1}$ is greater or equal to a certain quantity. However, since we do not observe any amount of dust grains in the region between the planets we expect it to be completely unstable for small particles. The condition needed to reach such situation is called max packing and it corresponds to take the two planets as close as possible to have a still stable system. Therefore, the max packing condition is satisfied by the equation
\begin{equation}
\label{two}
a_{p,2}-a_{p,1}=2\sqrt{3} \Bigl(\frac{2M_{p}}{3M_*} \Bigr)^{1/3}\Bigl(\frac{a_{p,1}+a_{p,2}}{2}\Bigr).
\end{equation}
The other two equations that we need are the ones of Morrison \& Malhotra, \eqref{Mo1} and \eqref{Mo2}, from which we obtain $a_{p,1}$ and $a_{p,2}$ in the form
\begin{equation}
a_{p,1}=\frac{d_1}{1-1.2\mu^{0.28}}
\end{equation}
\begin{equation}
a_{p,2}=\frac{d_2}{1+1.7\mu^{0.31}}
\end{equation}
and substituting in \eqref{two} we get
\begin{equation}
\begin{split}
d_2-d_1&=\sqrt{3} \Bigl(\frac{2}{3}\Bigr)^{1/3}\mu^{1/3}(d_1+d_2)+\\
               &+\sqrt{3} \Bigl(\frac{2}{3}\Bigr)^{1/3}(d_1 1.7\mu^{0.31+1/3}-d_2 1.2 \mu^{0.28+1/3})+\\
               &+1.2 d_2 \mu^{0.28}+d_1 1.7 \mu^{0.31}.
\end{split}
\end{equation}
\normalsize
This is a very complex equation to solve for $M_p$ and we need to make some simplifications. We note that all the exponents of $\mu$ have very similar values with the exception of the two $\mu$ in the third term on the right side of the equation in which, however, the exponents are about double of all others. Thus, we choose as a mean value $\mu^{0.31}$ and in the third term $\mu^{0.62}$ for both terms in the brackets. Calling $x=\mu^{0.31}$ we have now to solve the quadratic equation
\begin{equation}
\begin{split}
 &\sqrt{3} \Bigl(\frac{2}{3}\Bigr)^{1/3}(1.2d_2-1.7d_1)x^2-\\
 &-\Bigl(1.2d_2+1.7d_1+ \sqrt{3} \Bigl(\frac{2}{3}\Bigr)^{1/3}(d_1+d_2)\Bigr)x+d_2-d_1=0.
 \end{split}
 \end{equation}
 \normalsize
 We can finally obtain the value of $M_p$, given the positions of the two belts and the mass of the star
 \begin{equation}
 \begin{split}
 M_p=M_*&\Biggl(\frac{1.2d_2+1.7d_1+ \sqrt{3} \Bigl(\frac{2}{3}\Bigr)^{1/3}(d_1+d_2)}{2 \sqrt{3} \Bigl(\frac{2}{3}\Bigr)^{1/3}(1.2d_2-1.7d_1)}-\\
           -& \frac{\sqrt{\Bigl(1.2d_2+1.7d_1+ \sqrt{3} \Bigl(\frac{2}{3}\Bigr)^{1/3}(d_2+d_1)\Bigr)^2-}}{2 \sqrt{3} \Bigl(\frac{2}{3}\Bigr)^{1/3}(1.2d_2-1.7d_1)}\\
           & \quad \frac{\overline{-4\sqrt{3} \Bigl(\frac{2}{3}\Bigr)^{1/3}(1.2d_2-1.7d_1)(d_2-d_1)}}{2 \sqrt{3} \Bigl(\frac{2}{3}\Bigr)^{1/3}(1.2d_2-1.7d_1)}\Biggr)^{10/31}.
 \end{split}          
 \end{equation}
 
\begin{figure}[h!]
\centering
\includegraphics[scale=0.3]{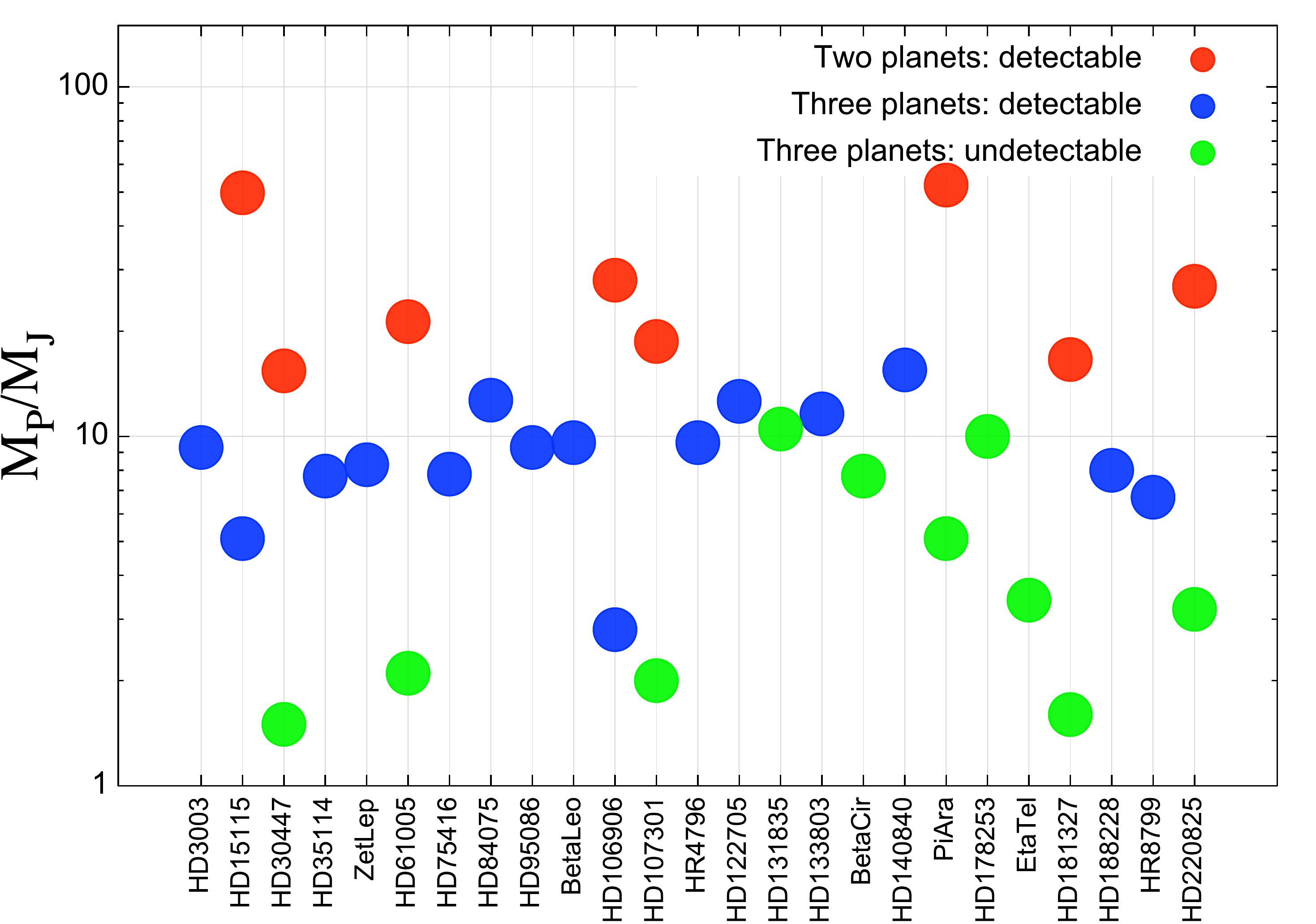}
\caption{Masses ($M_p/M_J$) for the systems with two and three equal-mass planets on circular orbits: red circles represent a system with two planets that are detectable, while green and blue circles represent three planets detectable and undetectable respectively. Since the equations are not fully correct for more massive planets, for the two planets-case we show only systems for which $M_P\le50M_J$ whereas for the three planets-case we show only systems for which $M_P\le15M_J$}
\label{23}
\end{figure}

The numerical outcomes of the equation show that this formula is reliable. We recall that our equations are valid only for $\mu\ll1$, thus we choose again as upper limit $50$ $M_J$ and arbitrarily we consider only masses bigger than $0.1$ $M_J$. In the case of two equal mass planets on coplanar circular orbits we obtain satisfying results only in 8 cases out of 35 presented in Table \ref{tabu2}.\\
The case of three planets of equal mass on circular orbits is quite similar and of particular interest.  Indeed, systems of three (or more) lower mass planets may be more likely sculptors than two massive planets on eccentric orbits that will be considered in the next Section, both because the occurrence rate of lower mass planets is higher than Jovian planets (at least in regions close to the star) as seen, for example, from Kepler \citep{Howard} or RV \citep{Mayor,Raymond} planets occurrence rates, and because the disk would not have to survive planet--planet scattering without being depleted \citep{Marzari1}.\\
For the three planets case, we have to consider four zones of instability for the particles: the first and the fourth are determined by the inner and the outer planet assuming equations \eqref{Mo1} and \eqref{Mo2} respectively, while the second and the third by the Hill criterion.\\
 From equation \eqref{semi}, we can express the mutual dependence between the positions of the three planets as
 \begin{equation}
 \label{H1}
 a_{p,2}=a_{p,1}+K\biggl(\frac{2M_p}{M_*}\biggr)^{1/3}\frac{a_{p,1}+a_{p,2}}{2}
 \end{equation}
 \begin{equation}
 \label{H2}
 a_{p,3}=a_{p,2}+K\biggl(\frac{2M_p}{M_*}\biggr)^{1/3}\frac{a_{p,2}+a_{p,3}}{2}.
 \end{equation}
We can obtain $a_{p,2}$ from equation \eqref{H1} and substituting it in \eqref{H2} we get
\begin{equation}
a_{p,3}=a_{p,1}\frac{\Bigl(1+\frac{K}{2}\Bigl(\frac{2}{3}\mu\Bigr)^{1/3}\Bigr)^2}{\Bigl(1-\frac{K}{2}\Bigl(\frac{2}{3}\mu\Bigr)^{1/3}\Bigr)^2},
\end{equation}
where $a_{p,1}$ and $a_{p,3}$ are determined by equations \eqref{Mo1} and \eqref{Mo2}. The final expression to solve for $M_p$ becomes
\begin{equation}
\frac{d_2}{d_1}\frac{1-1.2\mu^{0.28}}{1+1.7\mu^{0.31}}= \frac{\Bigl(1+\frac{K}{2}\Bigl(\frac{2}{3}\mu\Bigr)^{1/3}\Bigr)^2}{\Bigl(1-\frac{K}{2}\Bigl(\frac{2}{3}\mu\Bigr)^{1/3}\Bigr)^2}.
\end{equation}
In analogy with the previous cases, we impose a lower limit on the mass at $0.1$ $M_J$ but we have a further constraint on the upper one since values of $K$ are valid only up to some Jupiter masses. Thus we take as upper limit for the three planets model $15$ $M_J$. The values of $K$ are the ones described in the previous paragraph, with $K=8$ for masses up to $0.3$ $M_J$, $K=7$ for masses in the range $[0.3,0.9]$$M_J$ and $K=6$ for $M_p\ge1$$M_J$. For three equal mass planets on circular coplanar orbits we obtain more encouraging results since with such configuration the gap could be explained in 25 cases out of 35.
Results of the analysis of two and three planets on circular orbits are shown in Figure \ref{23}.\\ 
Together with the values of the masses for each system suitable to host two and/or three planets on circular orbit, we indicate the detectability of such planets comparing their masses and semi-major axis with SPHERE detection limits. The condition for detectability in this case is that at least one object in the two or three planets model is above the detection limits curve. However, as mentioned in Section 4, inclination of the disk may affect the detectability of the putative planets due to projection effects. Indeed, objects that are labeled as detectable in Figure \ref{23} are always observable only if the disk is face-on. With increasing inclination, the chance to detect the planets decreases. Whereas for the two-planets case the probability of detecting at least the outer objects is usually really high due to the big masses obtained, for the three-planets case half of the systems labeled as detectable are indeed observable only $50 \%$ of the time (the worst case is for the outer putative planet of HD133863 which would be detectable only $37\%$ of the time). \\
With the exception of very few systems, such as for example HD174429, HR8799, HD206893 and HD95086, no giant planets or brown dwarfs have been discovered between the two belts in the systems of our selected sample using direct imaging techniques.  Thus, we expect that if planets are indeed present, they must remain undetectable with our observations. In all systems, with the exception of HD131835, two planets on circular orbits would have been detected, since large masses are required. The situation quite improves for the three planets model because many systems can be explained with planets that would remain undetected. Therefore, in most cases, the assumption of three equal mass planets on circular orbits is more suitable than the one with two planets with the same characteristics.\\
Obviously in this paragraph we have made very restrictive hypotheses: circular orbits and equal mass planetary systems. Varying these two assumptions would give many suitable combinations in order to explain what we do (or do not) observe.\\

\subsubsection{Two planets on eccentric orbits}
\begin{figure} [h!]
\centering
\includegraphics[width=0.5\textwidth]{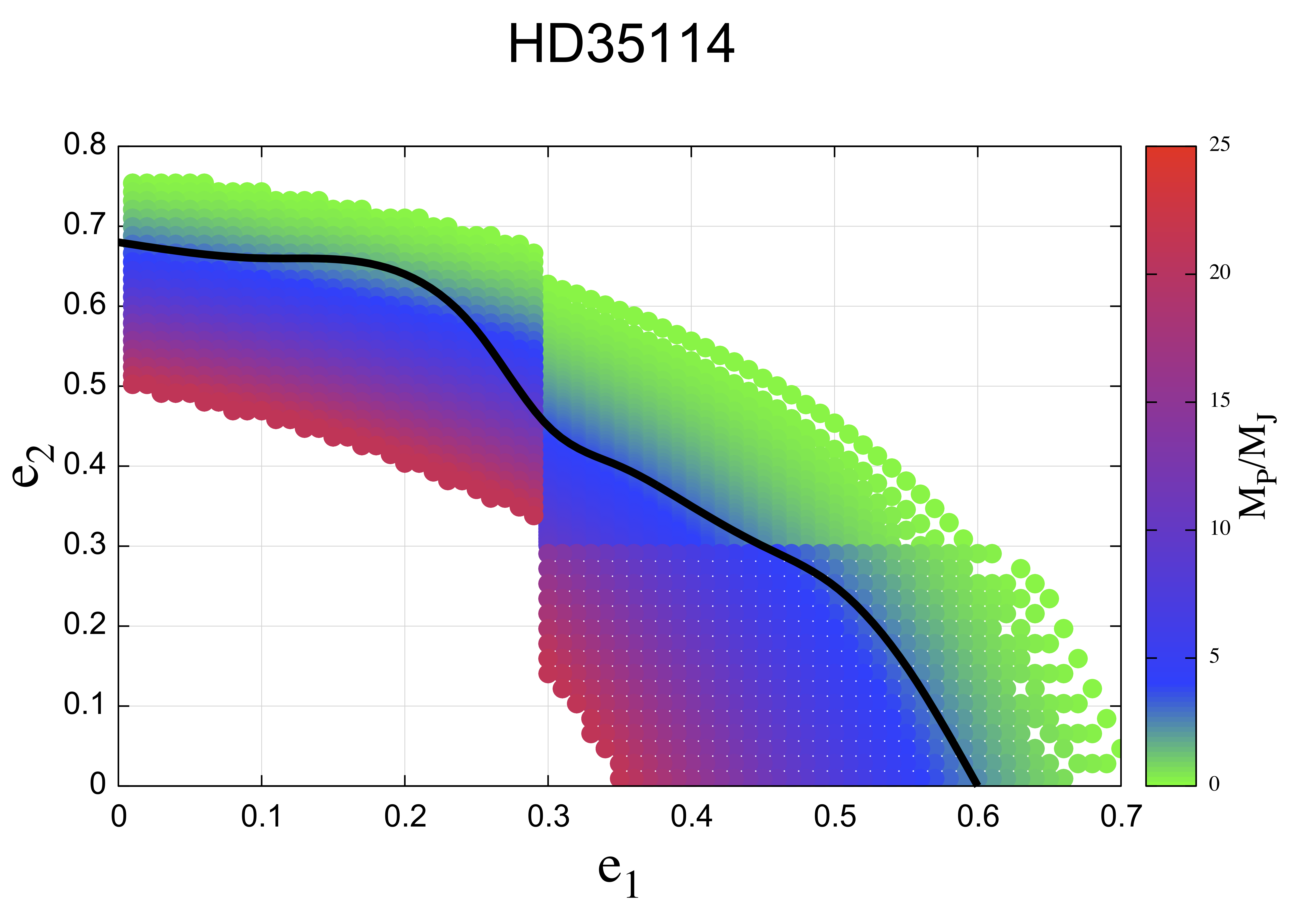}
\caption{Analysis for HD35114 with two equal-mass planets on eccentric orbits. On the axes the eccentricity of the inner ($e_1$) and outer ($e_2$) planets. The graduation of colors represent values of $M_p/M_J$. The black line represents the approximate detection limits: points above the line are undetectable whereas points below are detectable. The discontinuity at $e=0.3$ is due to the passage from Wisdom's equation to Mustill and Wyatt's one.}
\label{ecce}
\end{figure}

The last model we want to investigate is two equal-mass planets on eccentric orbits. The system is again divisible in three regions of stability. The zone between the two planets follows the Hill criterion for the condition of max packing given by equation \eqref{mass} with the equal sign. For the outer and inner regions the force is exerted by the planets on the massless bodies in the belts. This time, however, we will use Wisdom and Mustill \& Wyatt expressions instead of Morrison \& Malhotra's, suitable only for the circular case. Precisely, we apply the equation of Wisdom for eccentricities up to $0.3$ whereas for greater values of $e_p$ we use Mustill \& Wyatt, together with the substitution of $a_p$ with apoastron and periastron of the planets.\\
We have four different situations:
\begin{itemize}
\item if $e_{p,1}$ and $e_{p,2}$ are both $\le0.3$, then we used equations of Wisdom \eqref{apow} and \eqref{periw}, from which we obtain $a_{p,1}$ and $a_{p,2}$ in the form
\begin{equation}
a_{p,1}=\frac{d_1}{1-1.3\mu^{2/7}}\frac{1}{1-e_{p,1}}
\end{equation}
\begin{equation}
a_{p,2}=\frac{d_2}{1+1.3\mu^{2/7}}\frac{1}{1+e_{p,2}};
\end{equation}
\item if $e_{p,1}\le0.3$ and $e_{p,2}>0.3$ we apply at the inner planet the equation of Wisdom \eqref{periw} and at the outer one equation \eqref{apom} from Mustill \& Wyatt, thus obtaining
\begin{equation}
a_{p,1}=\frac{d_1}{1-1.3\mu^{2/7}}\frac{1}{1-e_{p,1}}
\end{equation}
\begin{equation}
a_{p,2}=\frac{d_2}{1+1.8\mu^{1/5}e_{p,2}^{1/5}}\frac{1}{1+e_{p,2}};
\end{equation}
\item if $e_{p,1}>0.3$ and $e_{p,2}\le0.3$ we have the opposite situation with respect to the one described above, thus we use Mustill \& Wyatt for the inner planet and Wisdom for the outer one
\begin{equation}
a_{p,1}=\frac{d_1}{1-1.8\mu^{1/5}e_{p,1}^{1/5}}\frac{1}{1-e_{p,1}};
\end{equation}
\begin{equation}
a_{p,2}=\frac{d_2}{1+1.3\mu^{2/7}}\frac{1}{1+e_{p,2}};
\end{equation}
\item if $e_{p,1}$ and $e_{p,2}$ are both $>0.3$ we use Mustill \& Wyatt for the two planets 
\begin{equation}
a_{p,1}=\frac{d_1}{1-1.8\mu^{1/5}e_{p,1}^{1/5}}\frac{1}{1-e_{p,1}};
\end{equation}
\begin{equation}
a_{p,2}=\frac{d_2}{1+1.8\mu^{1/5}e_{p,2}^{1/5}}\frac{1}{1+e_{p,2}};
\end{equation}
\end{itemize}
Thus, depending on the values of $e_{p,1}$ and $e_{p,2}$ we substitute in equation \eqref{mass} expressions of $a_{p,1}$ and $a_{p,2}$ as obtained above. Varying the masses in the range $[0.1,25]M_J$, we obtain the respective values of eccentricities for the two planets. We note that we are implicitly assuming that the two eccentric planets will remain on the same orbits for their whole lifetime whereas, in reality, their eccentricities will fluctuate. This may imply lower masses of the two planets required to dig the gap as the systems are unlikely to be observed at the peak of an eccentric cycle.\\
We show in Figure \ref{ecce} the results of this analysis for  HD35114. For each system, we obtain a set of suitable points identified by the three coordinates $[e_{p,1},e_{p,2},M_p]$ (we recall that the two planets in the system have the same mass). Therefore, we prepare a grid with the two values of eccentricities on the axes and we associate a scale of colors to the mass range (see Figure \ref{ecce} top panel). Moreover, in order to determine which planets would have been detected we confront, as always, values of semi-major axis and mass with the detection limits curves and use as a criterion of detectability the condition in which at least one of the two planets is above the curve even just in partial zones of its orbit (see Figure \ref{ecce}).\\
In the graphics, we indicate with a black line the approximate detection limits: points above the line are undetectable whereas points below are detectable.
From Figure \ref{ecce} it is clearly visible how mass (and thus detectability) decreases with increasing eccentricities. Moreover, small variations of $e_{p,1}$ and/or $e_{p,2}$ cause a great damp in mass since, as already mentioned above, the stability depends very little on the mass of the two planets.\\
From this study emerges that the apparent lack of giant planets in the sample of systems analyzed can easily be explained by taking quite eccentric planets of moderate masses that lay beneath detection limits curve. Indeed, large eccentricities are common features of exoplanets \citep{Udry} and thus we have not to abandon the hypothesis that gaps between two planetesimals belts are dig by the presence of massive objects that surround the central star.   

\section{Particularly interesting systems}

\subsection{HD106906}
\begin{figure} [h!]
\centering
\includegraphics[width=0.45\textwidth]{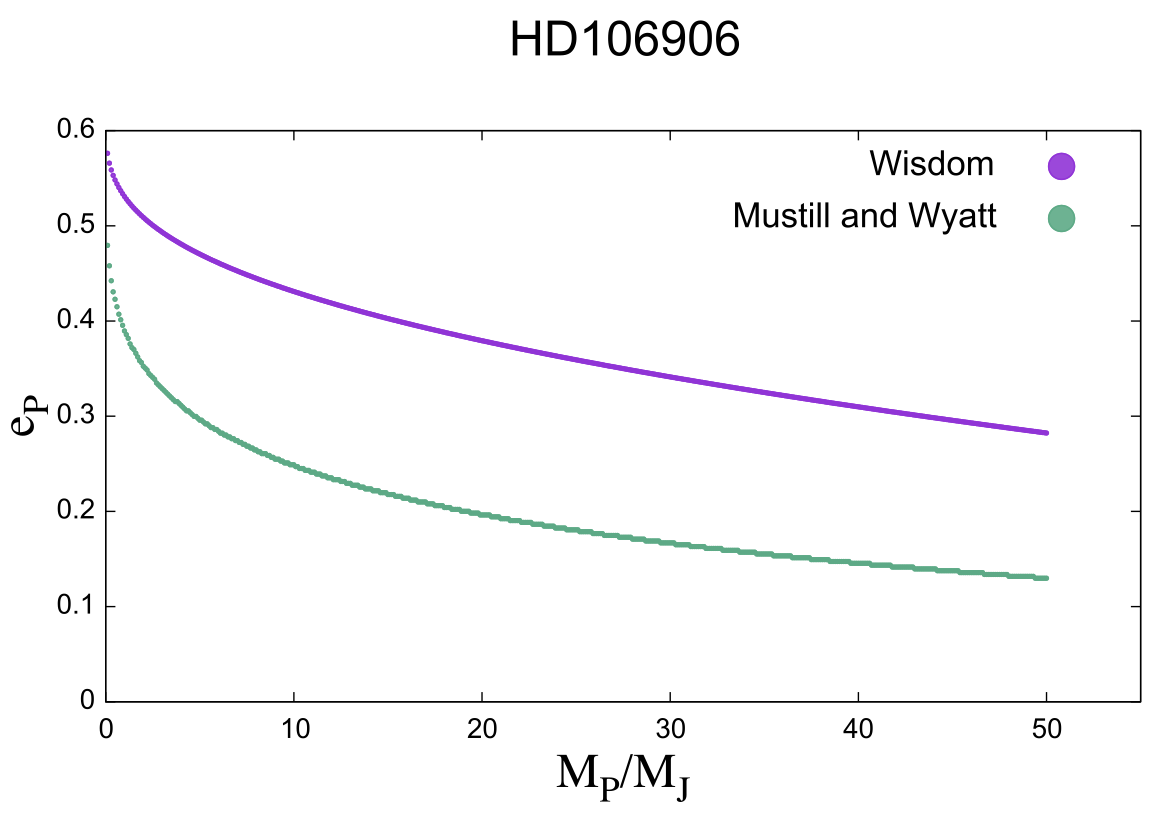}
\caption{One planet on eccentric orbit around HD106906.}
\label{HD106906_1}
\end{figure}

\begin{figure} [h!]
\centering
\includegraphics[width=0.5\textwidth]{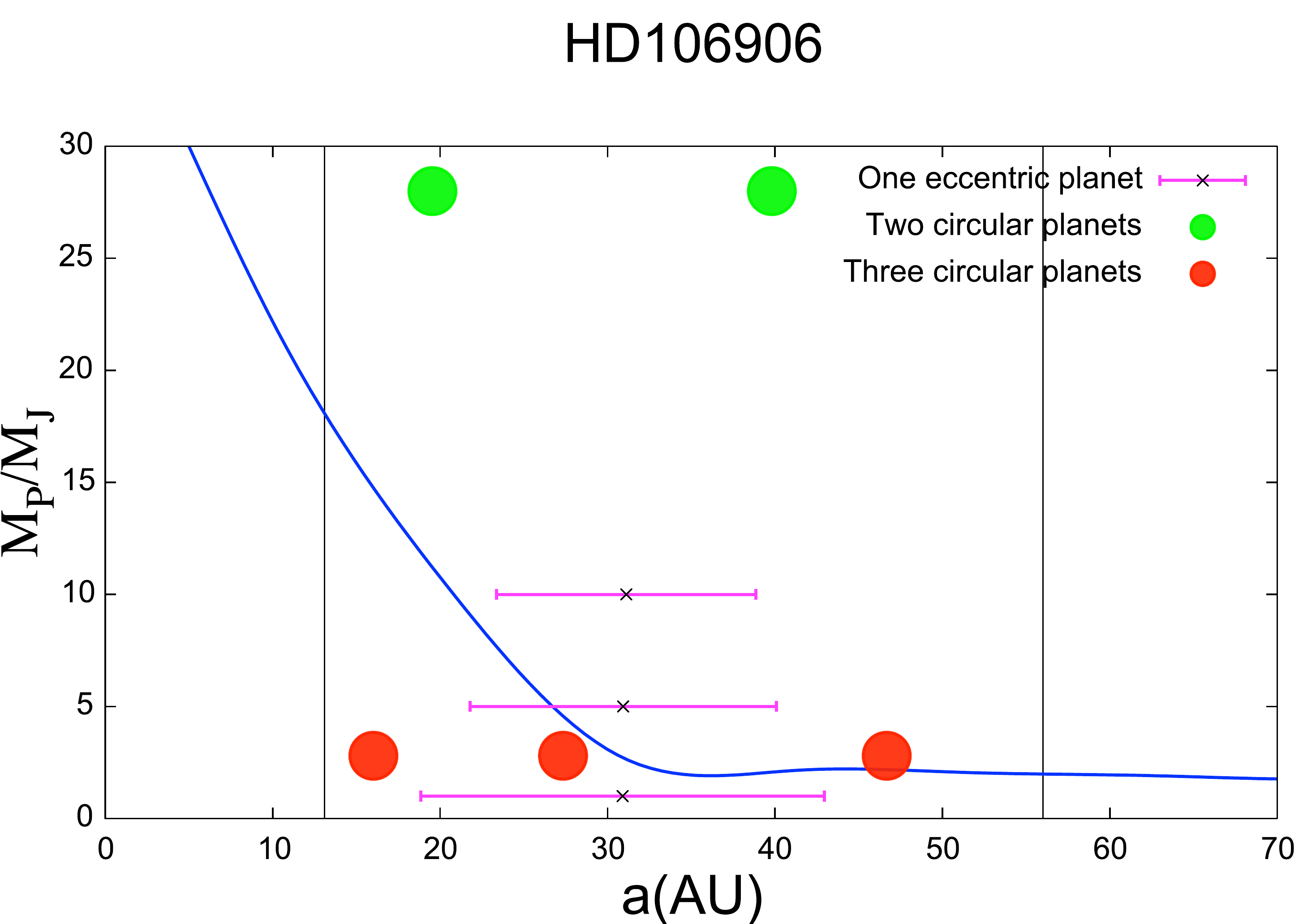}
\caption{Three different values of mass (1, 5 and 10 $M_J$) of putative planets with their respective semi-major axis and eccentricities and the detection limits curve for the HD106906 system (pink lines) and two (green circles) and three (red circles) planets on circular orbits around HD106906.}
\label{HD106906_1bis}
\end{figure}

\begin{figure} [h!]
\centering
\includegraphics[width=0.5\textwidth]{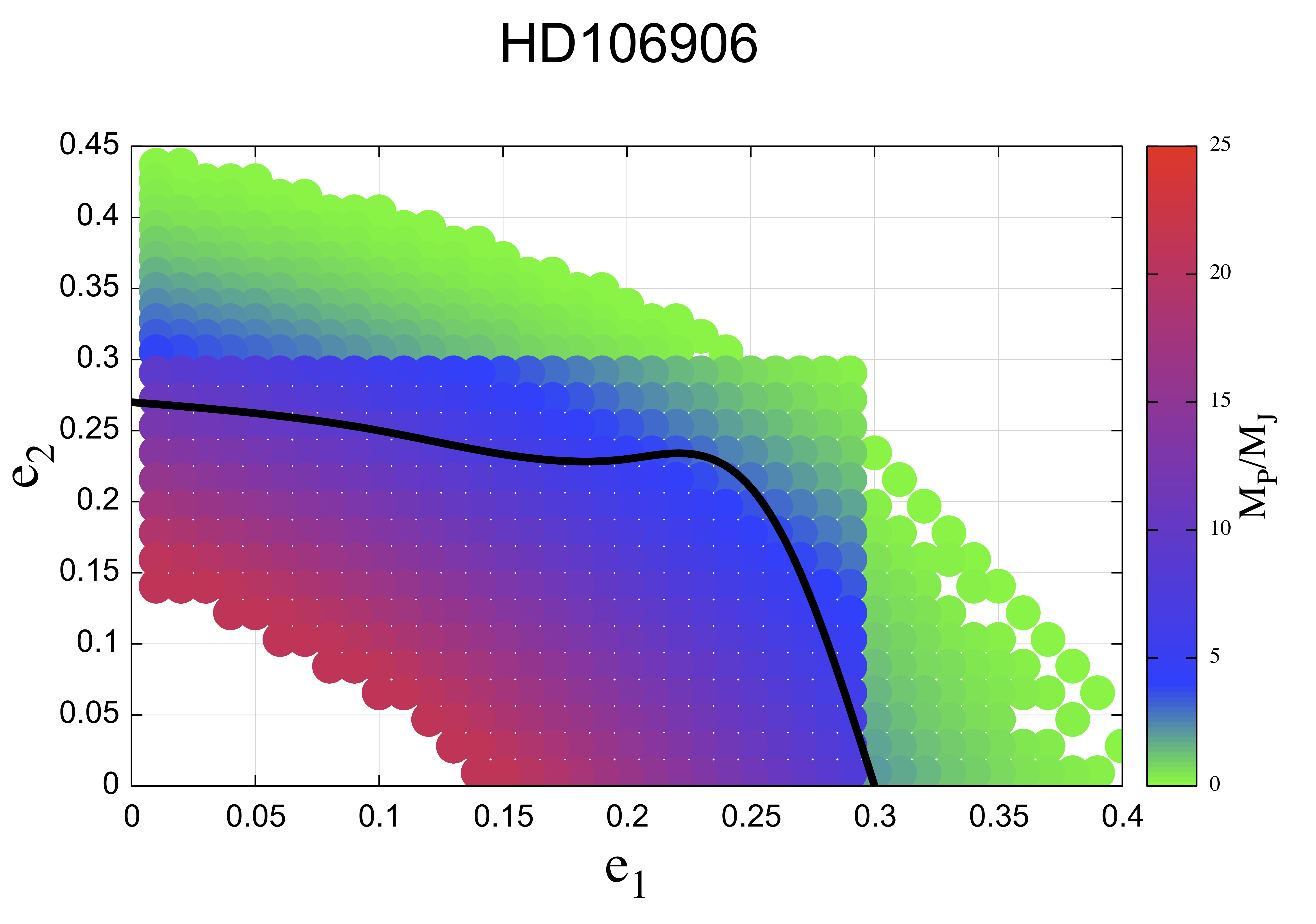}
\caption{Predictions and comparison with the SPHERE detection limits (black line) for the eccentric two-planet model for HD106906.}
\label{HD106906_3}
\end{figure}

HD106906AB is a close binary system (Lagrange et al. 2016, submitted) where both stars are F5 and are located at a distance of 91.8 pc. They belong to the Lower Centaurus Crux (LCC) group, which is a subgroup of the Scorpius–Centaurus (Sco-Cen) OB association. \cite{Bailey} detected a companion planet, HD106906 b, of  $11 \pm 2$  $M_J$ located at $\sim 650$ AU in projected separation and an asymmetric circumbinary debris disk nearly edge-on resolved by different instruments (see Appendix B). The evident asymmetries of the disk could be a hint of interactions between the planet and the disk \citep{Rodet, Nesvold}. The gap in the disk is located between 13.1 AU and 56 AU and the detected companion orbits far away from this area. Since the gap is quite small we find promising results for one or more undetected companions. As mentioned in Section 5, it is not possible to explain the gap with a single planet (with $M_P \le 50 M_J$) on circular orbit. In Figure \ref{HD106906_1} we show $e_{\mathrm{p}}$ vs. $M_{\mathrm{p}}$ for one eccentric planet to be responsible for the empty space between the belts: we do not need particularly high eccentricity even at low masses. Thus the gap, together with the non detection of another companion (besides HD106906 b), could be explained by a single planet on eccentric orbit as shown in Figure \ref{HD106906_1bis}. For example, a planet with a mass of 1 $M_J$, semi-major axis $a_p\sim 45$ AU and a reasonable value of eccentricity, $e_p\sim0.4$, would be able to dig the gap and be undetectable at the same time. However, we can consider a more complex architecture. In Figure \ref{HD106906_1bis} and \ref{HD106906_3} we show two and three equal-mass planets on circular orbits and two planets on eccentric orbits: whereas two planets on circular orbits would have been easily detected, two of the three circular planets-case are undetectable and the farthest one is very close to detection limits. Moreover, combining the orbital parameters of the third planets with the projection effects due to the inclination of the disk ($i=85^{\circ}$) we obtain that it would be detectable only $55\%$ of the time. Eventually, two planets on eccentric orbits with eccentricities $\ge 0.2$ would be under detection limits curve. 

\subsection{HD174429 (PZ Tel)}
HD174429, or PZ Tel A, is a G9IV star member of the moving group $\beta$ Pictoris. It is located at a distance of $49.7$ pc and a M brown dwarf companion was discovered independently by \cite{Mugrauer} and \cite{Biller} to orbit this star at a separation of $\sim25AU$ on a very eccentric orbit \citep[$e>0.66$,][] {Maire}. The mass of PZ Tel B varies in the range $[20,40]$ $M_J$ depending on the age of the system (\cite{Ginsky}; \cite{Schmidt}). \\
From SED fitting, \cite{Chen} found that the debris disk of PZ Tel is better represented by two components. However, excesses in the near-IR typical of the warm component were never observed and small deviation from the spectral energy distribution of the star could be attributed to the presence of the brown dwarf. For what regard the cold component, \cite{Chen} obtained a temperature of $39$ K that, together with the $\Gamma$ correction, places the belt at $317.5$ AU. In spite of this, \cite{Riviere} rejected the hypothesis of the presence of a debris disk because they did not find infrared excesses with Herschel/PACS at 70, 100 and 160 $\mu m$. \\
Since the literature on this disk is quite poor and discordant, we want to apply our method to the known companion in order to check if it can add constraints on the existence of the disk. The orbit of the brown dwarf is still a matter of debate but we know that it has to be very eccentric. Thus, we can use $30$ $M_J$ as a mean value for the mass and the three best orbits presented in Table 13 of \cite{Maire} together with the formulation for one single planet on eccentric orbit. We obtained as a result that for the first two orbits the planet does not cross the external disk placing the edge at $\sim250$ AU for the first one (very near to our estimated inner edge) and at $\sim150$ AU in the second one (leaving some free dynamical space). The third orbit, instead, would cross the disk and destroy its configuration. Thus, we cannot exclude completely the presence of a cold debris disk component.\\

\subsection{HR8799}
\begin{figure} [h!]
\centering
\includegraphics[width=0.45\textwidth]{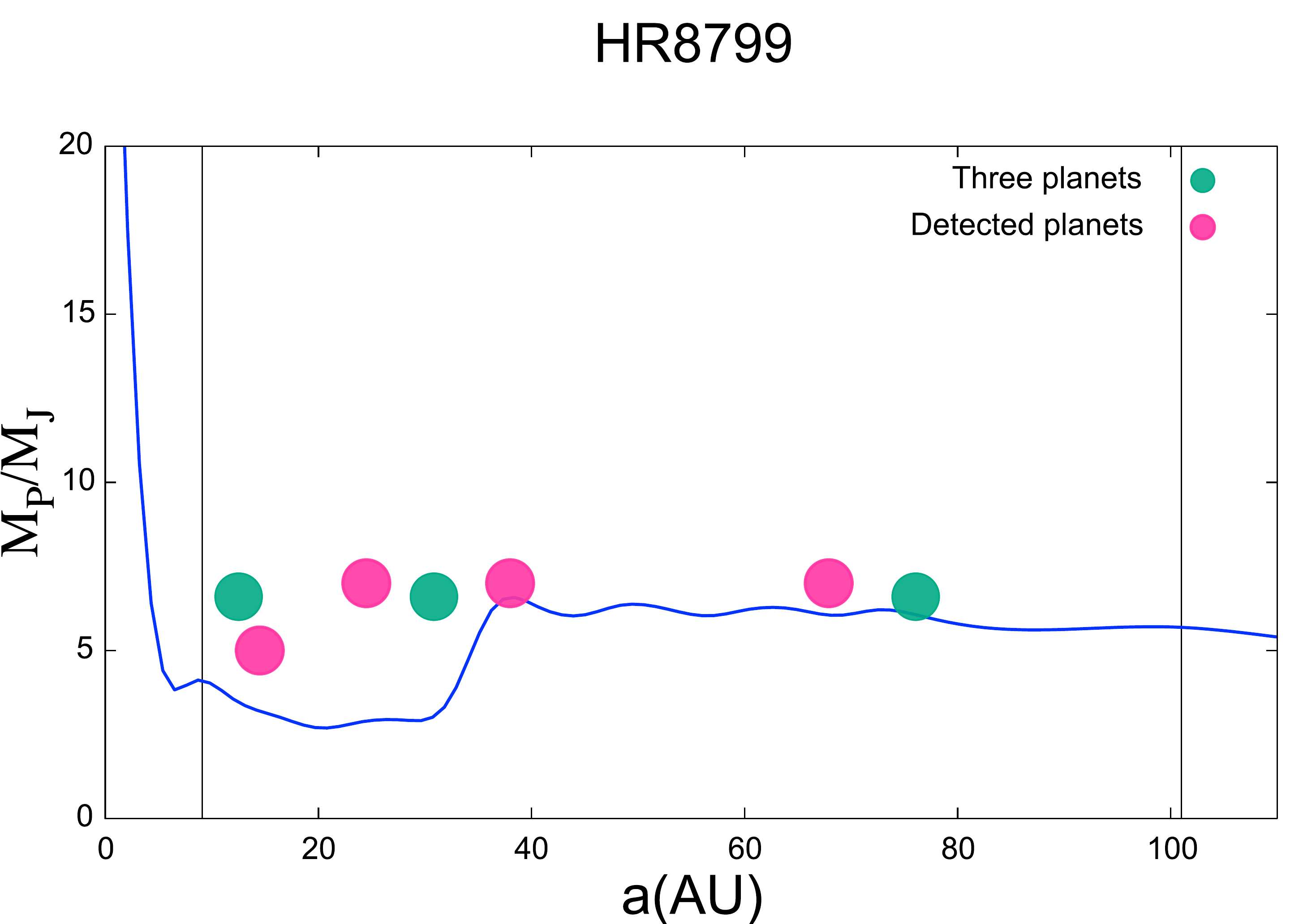}
\caption{Analytical results for three planets on circular orbits (green circles) and the actual four detected planets (pink circles) around HR8799 .}
\label{HR8799}
\end{figure}

HR8799 is a $\gamma$  Dor-type variable star \citep{Gray}. The most incredible characteristic of HR8799 is that it hosts four giant planets in the gap between the two components of the disk, with masses in the range $[5,7]$$M_J$ and distances in the range $[15,70]$ AU \citep{Marois}. Moreover, the disk around this star is spatially resolved in its outer component in far IR and millimeter wavelengths \citep{Su2, Booth2} and, besides the warm and cold belts placed at $\sim 9$ AU and $\sim 200$ AU respectively, it shows an extended halo up to $2000$ AU. 
Our dynamical analysis takes into account three planets so that we are not able to determine the precise dynamical behavior of HR8799 bcde. However, we can make some guess starting from our results. We show the results for three planets on circular orbits in Figure \ref{HR8799} that are the ones that comes nearer to the real architecture of HR8799 represented by the pink circles in the same figure \citep[][indeed, orbits with large eccentricities are not favored by current analyses]{Konopacky}. We find three equal-mass planets of $6.6$ $M_J$ that are similar to the estimated masses of HR8799 bcde. Thus it seems that this system, in order to be stable, must be dynamically full and even more packed to host four giant planets and mean motion resonances, that are not included in our analysis, may be at work \citep{Esposito, Gozdziewski}. 

\subsection{HR4796}

\begin{figure} [h!]
\centering
\includegraphics[width=0.45\textwidth]{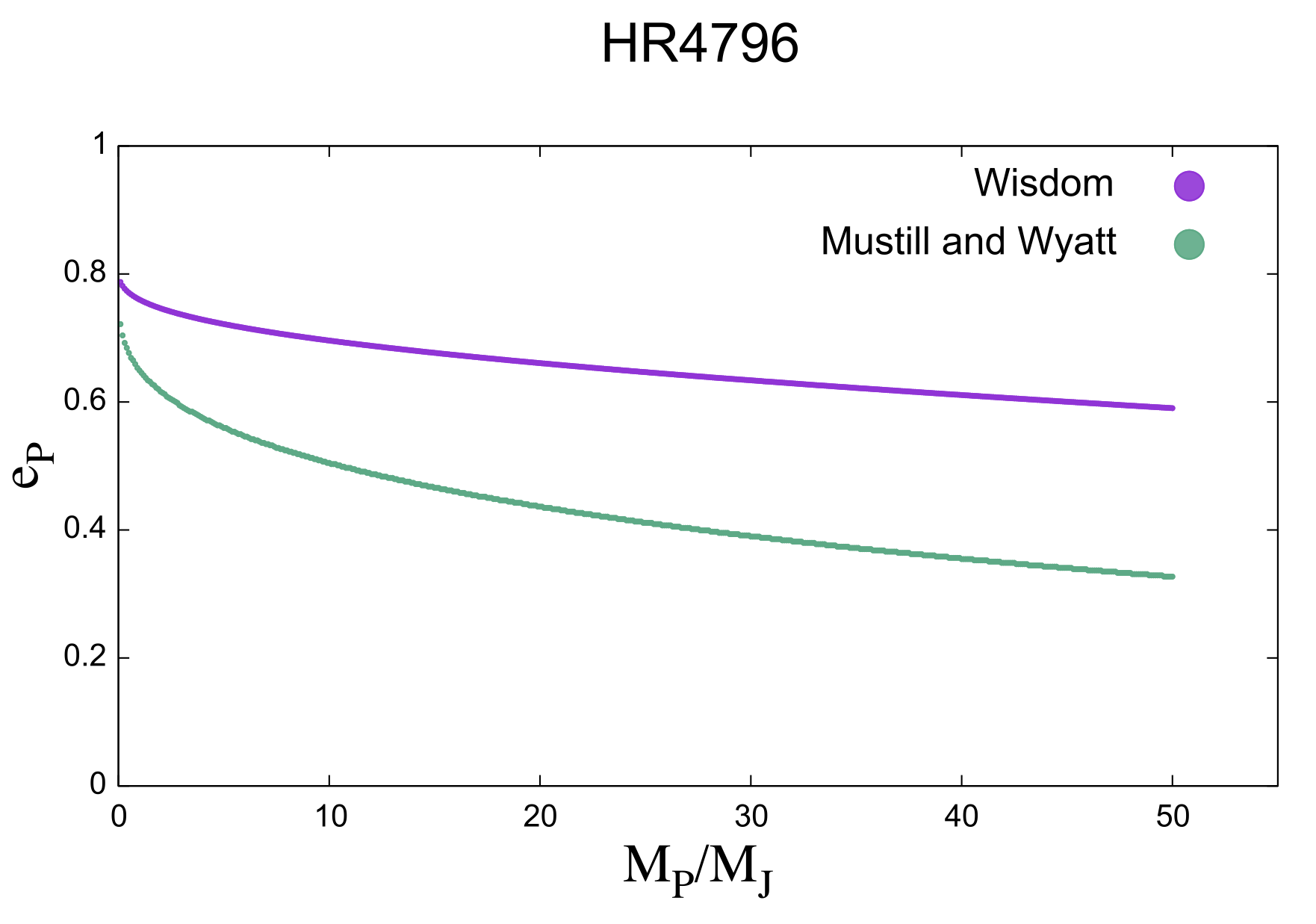}
\caption{One planet on eccentric orbit around HR4796.}
\label{HR4796_1}
\end{figure}

\begin{figure} [h!]
\centering
\includegraphics[width=0.5\textwidth]{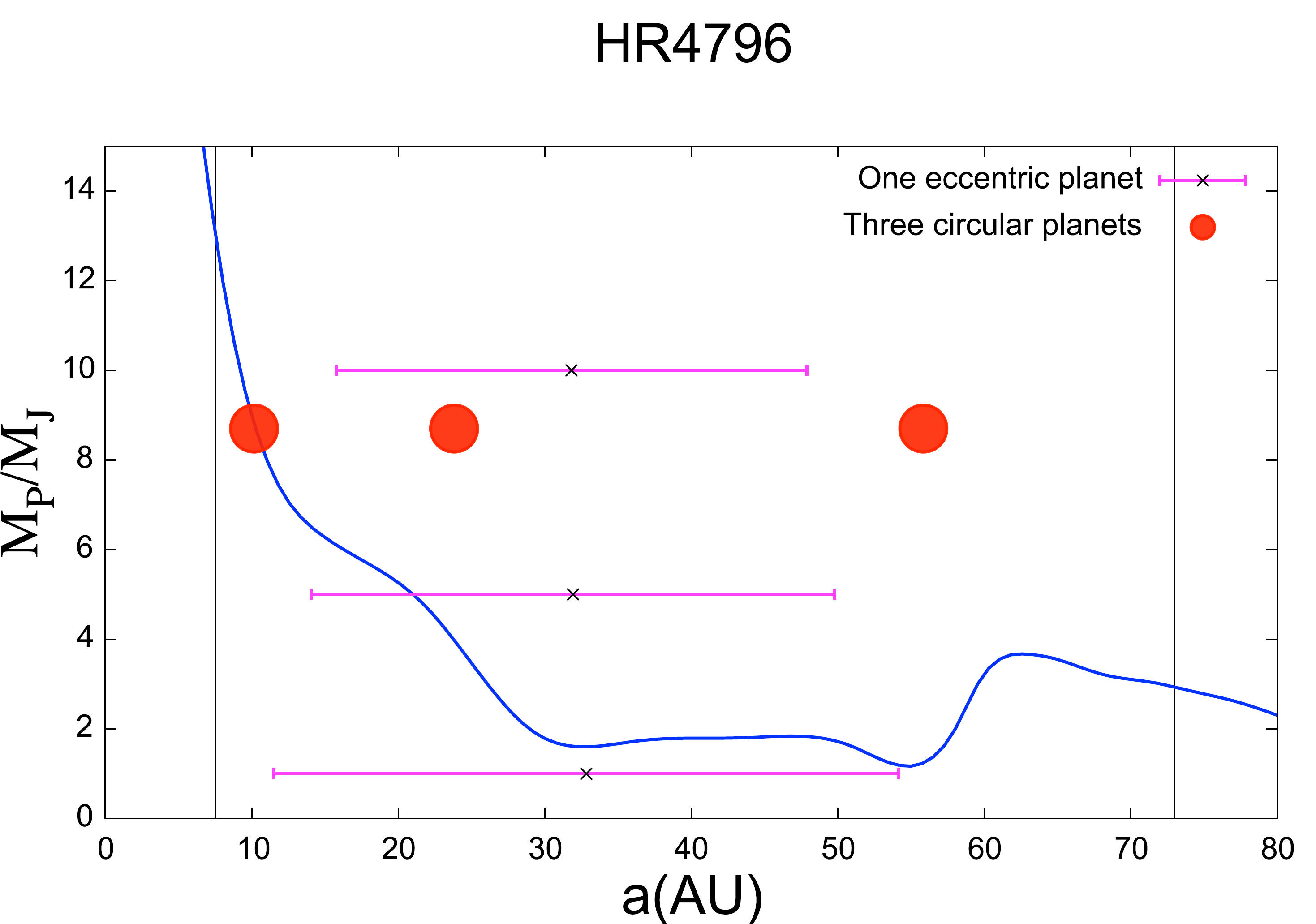}
\caption{Three different values of mass (1, 5 and 10 $M_J$) with their respective semi-major axis and eccentricities and the detection limits curve for the system (pink lines) and three planets on circular coplanar orbits around HR4796 (red circles).}
\label{HR4796_1bis}
\end{figure}

\begin{figure} [h!]
\centering
\includegraphics[width=0.5\textwidth]{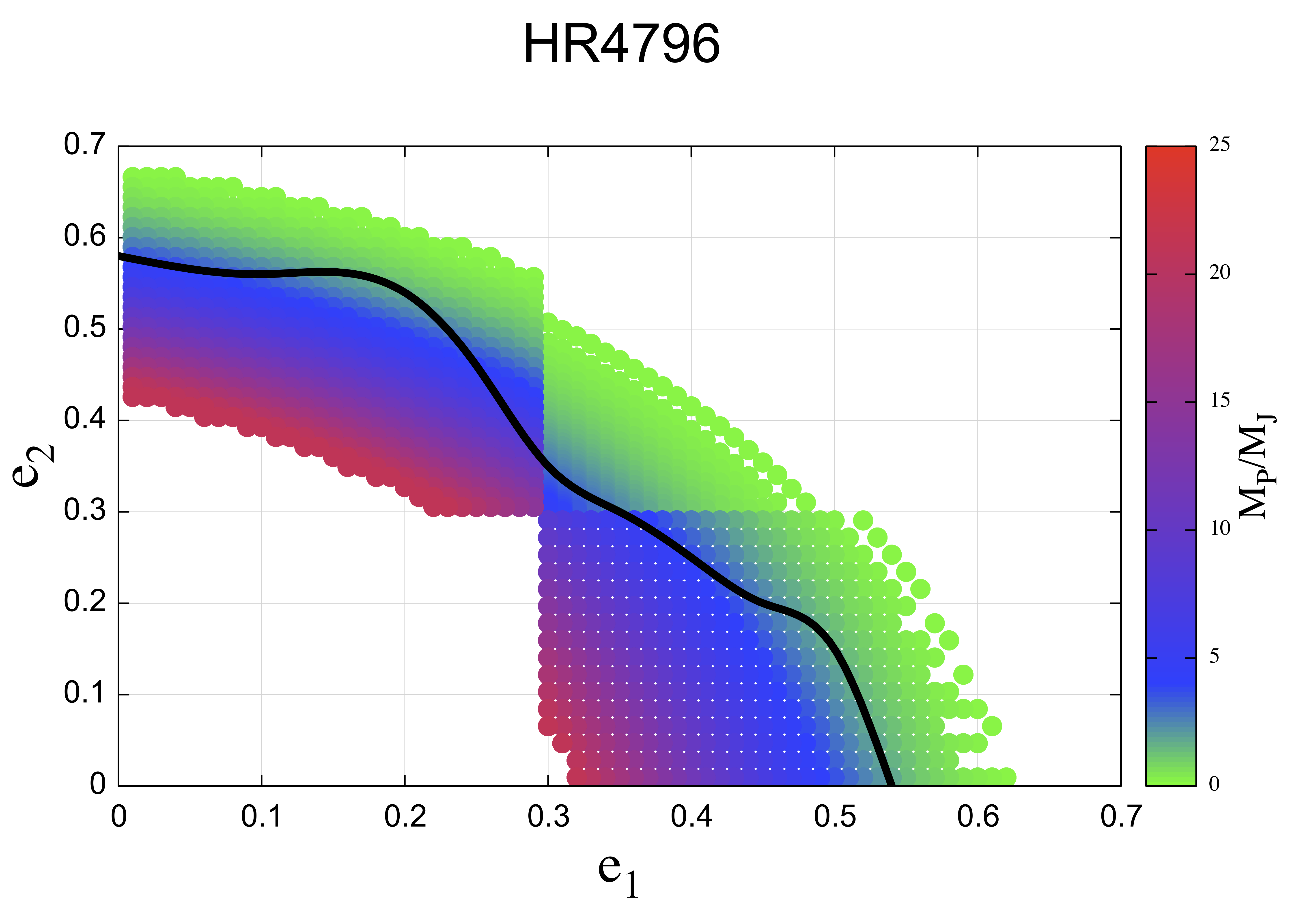}
\caption{Two planets on eccentric orbits around HR4796 and comparison with detection limits (black line).}
\label{HR4796_2}
\end{figure}

HR4796A is a A0 star and it is part of the TW Hydra kinematic group. It is part of a binary system with the M2 companion HR4796B orbiting at a projected separation of 560 AU. The debris disk around this star is resolved in scattered light and near IR. The images show a thin, highly inclined ring with an eccentricity of $\sim 0.06$ at $\sim 75$ AU from the star \citep{Kastner}. The eccentricity of the disk and the sharpness of its inner edge seems to point to the presence of a planet orbiting right inside the gap. However, no companion was detected so far \citep{Milli2}. Thus we would expect a planet with small mass under detection limits curve or with high eccentricity such that it passes, at some time of its orbit, in areas sufficiently near to the star even if the second hypothesis should imply an higher forced eccentricity of the disk. Comparing Figure \ref{HR4796_1} and \ref{HR4796_1bis} we obtain that objects with masses $\le 2$ $M_J$ are undetectable but, at the same time, they need high eccentricities ($\ge0.6$). Moreover, if we consider equation \eqref{ee} for the forced eccentricity exerted by the planet on the belt we should expect an eccentricity of the latter of $\sim 0.1$ that would require masses $\ge 50$ $M_J$ well above detection limits curve. Thus, we consider a more complex configuration with two and three planets (Figures \ref{HR4796_1bis} and \ref{HR4796_2}). No result was found for two planets on circular orbits whereas for three companions we find masses of $8.7$ $M_J$ each. In this last scenario two of the three planets would have been detected (even taking into account the inclination of the disk, $i=75^{\circ}$, the farthest putative planet would be always detectable) so we move to consider the case of two planets on eccentric orbits. Since, as explained in Section 6, the eccentricities of the two planets have a greater importance than the masses for this kind of analysis, we finally find possible solutions to explain the gap with the presence of undetectable objects (see the ellipses in Figure \ref{HR4796_2}).

\begin{table*} 
\caption{Values of masses and eccentricities needed to carve the gaps for the systems in the sample assuming one, two or three planets. We indicate with an asterisk undetectable objects. Since for the eccentric case of one single planet solutions are not univocally determined, in columns 3 we show only ($M_P,e_P$) at the boundary between detectability and undetectability. We adopt the same criterion in column 5 for two equal-mass planets on eccentric orbits. However, for this last case we obtained, given a fixed value of the mass, different possible combinations of the eccentricities of the two planets. Thus, for each system, we show only the combination for which values for $e_1$ and $e_2$ are similar.  }
\label{conclusion}
\centering
\begin{tabular}{ccccccccc}
\hline\hline
Name & \multicolumn{2}{c}{One Planet}&&\multicolumn{3}{c}{Two Planets}&&Three Planets\\
           &        Circular&\multicolumn{2}{c}{Eccentric}&Circular&\multicolumn{3}{c}{Eccentric}&Circular\\
           &      $M_P/M_J$      & $M_P/M_J$&$e_P$&$M_P/M_J$& $M_P/M_J$ &$e_1$&$e_2$&$M_P/M_J$\\
\hline  
HD1466  &  $\ge50$  &  $2.7$  &  0.89  &  $\ge50$  & 2.7  & 0.59  & 0.61  &  $\ge15$  \\
HD3003  &  $\ge50$  &  $1.7$  &  0.66  &  $\ge50$  & 1.7 & 0.36  & 0.36  &  $9.3$      \\
HD15115  &  $\ge50$  &  3.5  & 0.55  & $49.8$  & 3.5 & 0.33  & 0.29  & $5.1$\\
HD30447  &  $\ge50$  &  5.9  &  0.24 &  $15.4$ & 5.9  & 0.15 & 0.16 & $1.5$$^*$\\ 
HD35114  &  $\ge50$  &  2.2  &  0.69 &  $\ge50$ & 2.2 & 0.39 & 0.39 & $7.7$\\
$\zeta$ Lep & $\ge50$  & 6.5  & 0.56  & $\ge50$  & 6.5 & 0.34 & 0.29 & $8.3$\\
HD43989    & $\ge50$  & 2.8 & 0.93  &$\ge50$  & 2.8 & 0.65 & 0.65 & $\ge15$\\
HD61005  & $\ge50$ & 2.7 & 0.42 & $21.3$ & 2.7 & 0.29 & 0.29 &  $2.1$$^*$\\
HD71155  &  $\ge50$ &  11.2 & 0.80 & $\ge50$  & 11.2 & 0.48 & 0.48 & $\ge15$\\
HD75416  &  $\ge50$ &  4.3 & 0.50 & $\ge50$ & 4.3 & 0.30 & 0.27 & $7.8$\\
HD84075  &  $\ge50$ &  2.3 & 0.83 & $\ge50$ & 2.3 & 0.51 & 0.51 & $12.7$\\
HD95086  &  $\ge50$ & 2.0 & 0.70 & $\ge50$ & 2.0 & 0.39 & 0.39 & $9.3$\\
$\beta$ Leo  &  $\ge50$  & 1.6 & 0.68 & $\ge50$ & 1.6 & 0.38 & 0.39 & $9.6$\\
HD106906  &  $\ge50$  & 1.9 & 0.36 &  $28.0$  & 1.9 & 0.24 & 0.25 &  $2.8$\\
HD107301  &  $\ge50$  &  5.0 & 0.27 &  $18.7$ & 5.0 & 0.18 & 0.20 & $2.0$$^*$\\
HR4796      &  $\ge50$  & 1.2 & 0.64 &  $\ge50$  & 1.2 & 0.35 & 0.35 &  $9.6$\\
$\rho$ Vir &  $\ge50$  & 6.0 & 0.89 &  $\ge50$  & 6.0 & 0.59 & 0.59 & $\ge15$\\
HD122705  &  $\ge50$  & 7.0 & 0.66  &  $\ge50$  & 7.0 & 0.36 & 0.36 & $12.6$\\
HD131835  &  $\ge50$  & 4.0 & 0.65  & $\ge50$  & 4.0 & 0.36 & 0.35 & $10.5$$^*$\\
HD133803  &  $\ge50$  &  10.0 &  0.61  &  $\ge50$  & 10.0 & 0.32 & 0.33 &  $11.6$  \\
$\beta$ Cir  &  $\ge50$  &  13.4  &  0.45  &  $\ge50$  & 13.4 & 0.26 & 0.30 &  $7.7$$^*$\\
HD140840  &  $\ge50$  &   5.4  &  0.70  &   $\ge50$  & 5.4 & 0.39 & 0.40 &  $15$\\
HD141378  &  $\ge50$  &   9.0  &  0.86  &  $\ge50$  & 9.0 & 0.54  & 0.55 &  $\ge15$\\
$\pi$ Ara    &   $\ge50$  &   11.8 & 0.48 &  $50$  & 11.8 & 0.29 & 0.33 & $5.1$$^*$\\
HD174429  &  $\ge50$  &   2.4  &  0.99  &  $\ge50$  & 2.4 & 0.83 & 0.82 &  $\ge15$\\
HD178253  &  $\ge50$  &  9.9  &  0.55  &  $\ge50$  & 9.9 & 0.32 & 0.29 &  $10$$^*$\\
$\eta$ Tel   &  $\ge50$  &   3.8  & 0.36 &   $30.2$  & 3.8 & 0.25 & 0.25 & $3.4$$^*$\\
HD181327  &  $\ge50$  &  4.3  &  0.28  &  $16.6$  & 4.3 & 0.04 & 0.29 &  $1.6$$^*$\\
HD188228  &  $\ge50$  &  3.3  &  0.57  &  $\ge50$  & 3.3 & 0.30 & 0.31 & $8.0$\\
$\rho$ Aql   &  $\ge50$  &  10.1& 0.84 &  $\ge50$  & 10.1 & 0.52 & 0.52 &  $\ge15$\\
HD202917  &  $\ge50$  &  8.7  & 0.87 &  $\ge50$  & 8.7 & 0.56 & 0.56 &  $\ge15$\\
HD206893  &  $\ge50$  &  4.2  &  0.93  &  $\ge50$  & 4.2 & 0.66 &  0.67 & $\ge15$\\
HR8799      &  $\ge50$  &  2.7  &  0.61  &  $\ge50$  & 2.7 & 0.33 & 0.33 & $6.7$\\
HD219482  &  $\ge50$  &  7.5  &  0.90  &  $\ge50$  & 7.5 & 0.60 & 0.61 &  $\ge15$\\
HD220825  &  $\ge50$  &  12.0  & 0.30 &  $26.9$  & 12.0 & 0.2 & 0.2 & $3.2$$^*$\\
\hline  
\end{tabular}
\end{table*}

\section{Conclusions and perspectives}
In this work we studied systems that harbor two debris belts and a gap between them. The main assumption was that one or more planets are responsible for the gap. In a sample of 35 systems with double belts also observed as  part of the SPHERE GTO survey we found no planet or brown dwarf within the gap with the exception of HD218396 (HR8799), HD174429 (PZ Tel), HD206893 and HD95086. We note that some systems in our sample have detected and/or candidate companions that however orbit outside the gap (Langlois et al. 2017, in prep). The lack of planet detections within the belt may be due either to the dynamical and physical properties of the planets placing them below the detection limits of actual instruments or to some complex mechanism for which such systems were born with two separate disk components \citep{Kral2} .\\
We focus on the first hypothesis and test the detectability of different packed planetary systems which may carve the observed gap. We first investigated the presence of one single planet on a circular orbit and find that  for the systems in our sample most planets should have been detected or are too massive to be termed planets.\\
Our next step was to consider a single planet in an eccentric orbit using the equations of  Wisdom or Mustill \& Wyatt for the chaotic zone of a planet on circular orbit but replacing in the equations the semi-major axis of the planet with its periastron or  apoastron to compute the border of the inner or outer belt, respectively. 
For larger eccentricities, the mass of the planet decreases slightly but  extreme parameters are still required to model the double belt structure. To have undetectable planets with masses beneath detection limits  we predict  values of $e_p$, greater than $0.7$ in most systems. Thus, even if the hypothesis of one planet could be suitable in some cases, we try a different approach assuming that the gap is created by a packed system of lower mass planets.
We have explored the case of two or three planets in the gap to see if we can explain the gap and the non--detection of planets by SPHERE. The first model considers two equal-mass planets on circular orbits but even in this case most of the planets should have been detected by  SPHERE. The most promising scenario models the presence of the gap as due to the perturbations of either a system of three planets on circular orbits or two planets on eccentric orbits. In both these scenarios the planets would be undetected by SPHERE. In particular, the case with eccentric orbits show that even a small variation in the eccentricity of one of the two planets lead to a drop in their masses  hiding them from possible detections.\\
We summarize our result in Table \ref{conclusion}. We show values of masses for the case of one, two and three planets on circular orbits as found by our dynamical analysis. For the case of three planets, we indicate with an asterisk the undetectable objects. We show also results for masses and eccentricities for one and two planets on eccentric orbits: in eccentric cases the orbital parameters are not univocally determined because of the degeneracy between the mass and the eccentricity of the planet. Thus, for one single planet on an eccentric orbit, we show only the ($M_P,e_P$) values at the boundary between detectability and undetectability. The same criterion is applied in discerning values of ($M_P,e_1, e_2$) for two planets on eccentric orbits. However, we have to apply a further selection in this case since, given a fixed value of the mass, different possible combinations of the eccentricities of the two planets are possible. Therefore, for each system, we choose to show only the combination for which values of $e_1$ and $e_2$ are similar. \\
We  conclude our work noticing that, even if very few planets have been detected so far in the gap of double belts, we cannot rule out the hypothesis that the gap is indeed due to massive objects orbiting within it. One or two planets on circular orbits would have been revealed for each system, in contrast with our observations, and more complex architectures should be taken into account. Multiple planet systems with eccentric orbits may be responsible for these belts architectures as in the case of HR8799 or the Solar System.\\
This paper presents a quick method to estimate the masses and eccentricities of the planets in these packed configuration to have a first glimpse to the possible architecture of the planetary system in the gap. Thanks to the analytical formulations we can easily obtain masses, semi-major axes and eccentricities of planets responsible for the gap. When the SPHERE GTO program will be completed, the number of observed two-components disks will be more than doubled. All these systems will be suitable for this kind of analysis and a statistical study on the presence of planets between belts will be possible. Moreover, spatially resolved images such as those that are and will be provided by instruments like SPHERE, GPI, ALMA and JWT, will be of extreme importance in order to get to more complete and robust conclusions on debris disks, exoplanets and their interactions.\\

\begin{acknowledgements}
SPHERE  is  an  instrument  designed and built by a consortium consisting of IPAG (Grenoble, France), MPIA (Heidelberg,  Germany), LAM  (Marseille,  France),  LESIA  (Paris,  France), Laboratoire Lagrange (Nice, France), INAF–Osservatorio di Padova (Italy), Observatoire  de  Geneve  (Switzerland),  ETH  Zurich  (Switzerland), NOVA  (Netherlands),  ONERA  (France)  and  ASTRON (Netherlands) in  collaboration  with ESO. SPHERE was funded by ESO, with additional contributions from CNRS (France),  MPIA  (Germany),  INAF  (Italy),  FINES  (Switzerland)  and  NOVA (Netherlands).  SPHERE  also  received  funding  from  the  European  
Commission  Sixth  and  Seventh  Framework  Programmes  as  part  of  the Optical  Infrared  Coordination  Network  for  Astronomy  (OPTICON)  under grant  number RII3-Ct-2004-001566 for FP6 (2004–2008), grant number 226604 for FP7 (2009–2012)  and  grant  number  312430  for  FP7  (2013–2016).  
This  work  has  made  use  of  the  the  SPHERE  Data Centre, jointly operated by OSUG/IPAG (Grenoble), PYTHEAS/LAM/CESAM (Marseille), OCA/Lagrange (Nice) and Observtoire de Paris/LESIA (Paris). We thank P.  Delorme  and  E.  Lagadec  (SPHERE  Data  Centre)  for  their  efficient help  during  the  data  reduction  process.  
We  acknowledge  financial  support from the “Progetti Premiali” funding scheme of the Italian Ministry of Education, University, and Research. We  acknowledge  financial  support  from  the  Programme National de Planetologie (PNP) and the Programme National de Physique Stellaire (PNPS) of CNRS-INSU. This work has also been supported by a grant from the French Labex OSUG\@2020 (Investissements d’avenir – ANR10 LABX56). The project is supported by CNRS, by the Agence Nationale de la Recherche
(ANR-14-CE33-0018;  ANR-16-CE31-0013).  Quentin Kral acknowledges funding from STFC via the Institute of Astronomy, Cambridge, Consolidated Grant.\\
We thank the referee Dr. A. Mustill for the useful comments.
\end{acknowledgements}

\bibliographystyle{aa}
\bibliography{bibliography}

\begin{appendix}

\section{$\Gamma$ correction}
\begin{figure} 
\begin{tabular}{@{}c@{}} 
\includegraphics[width=0.45\textwidth]{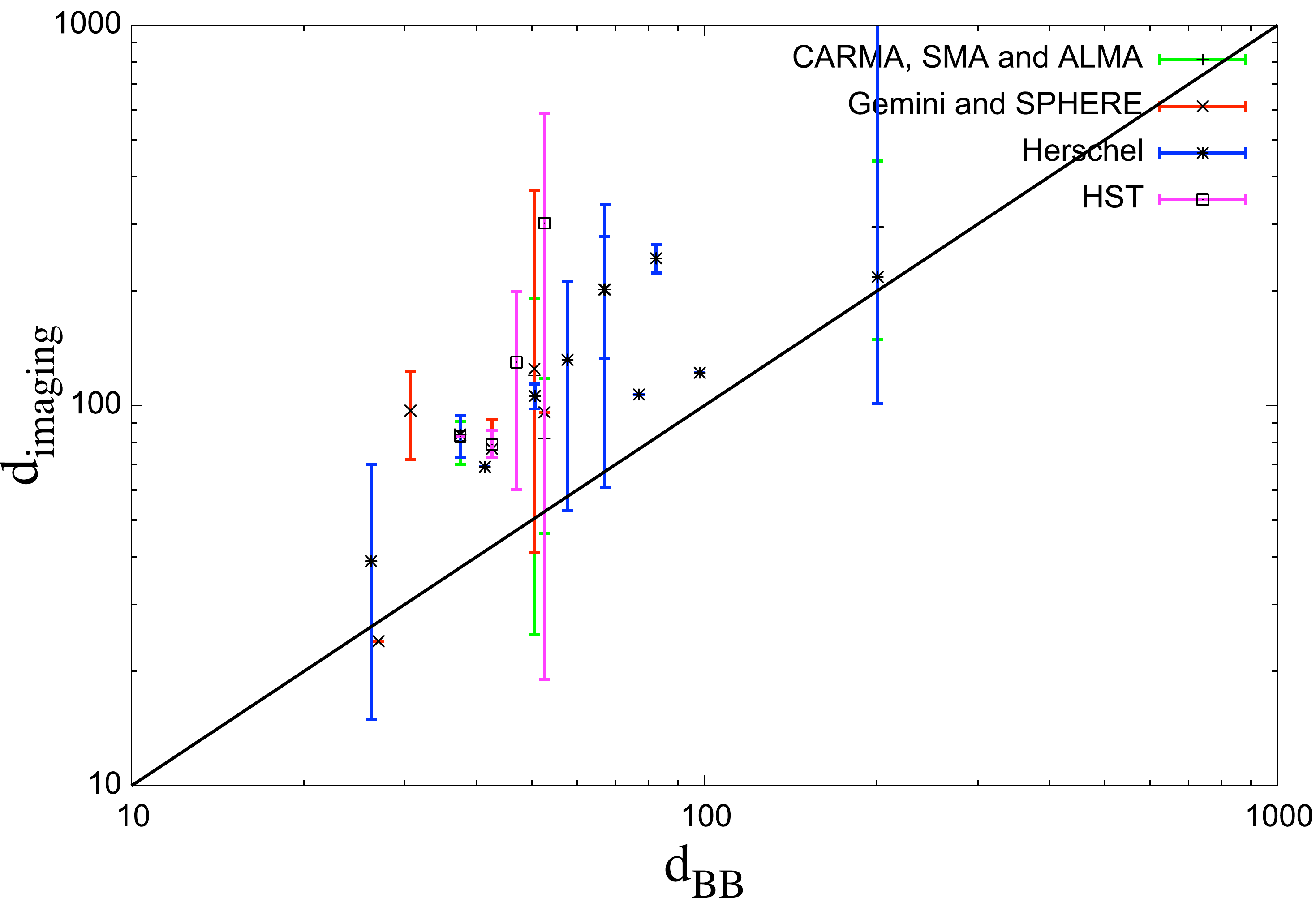}\\
\end{tabular}
\begin{tabular}{@{}c@{}} 
\includegraphics[width=0.45\textwidth]{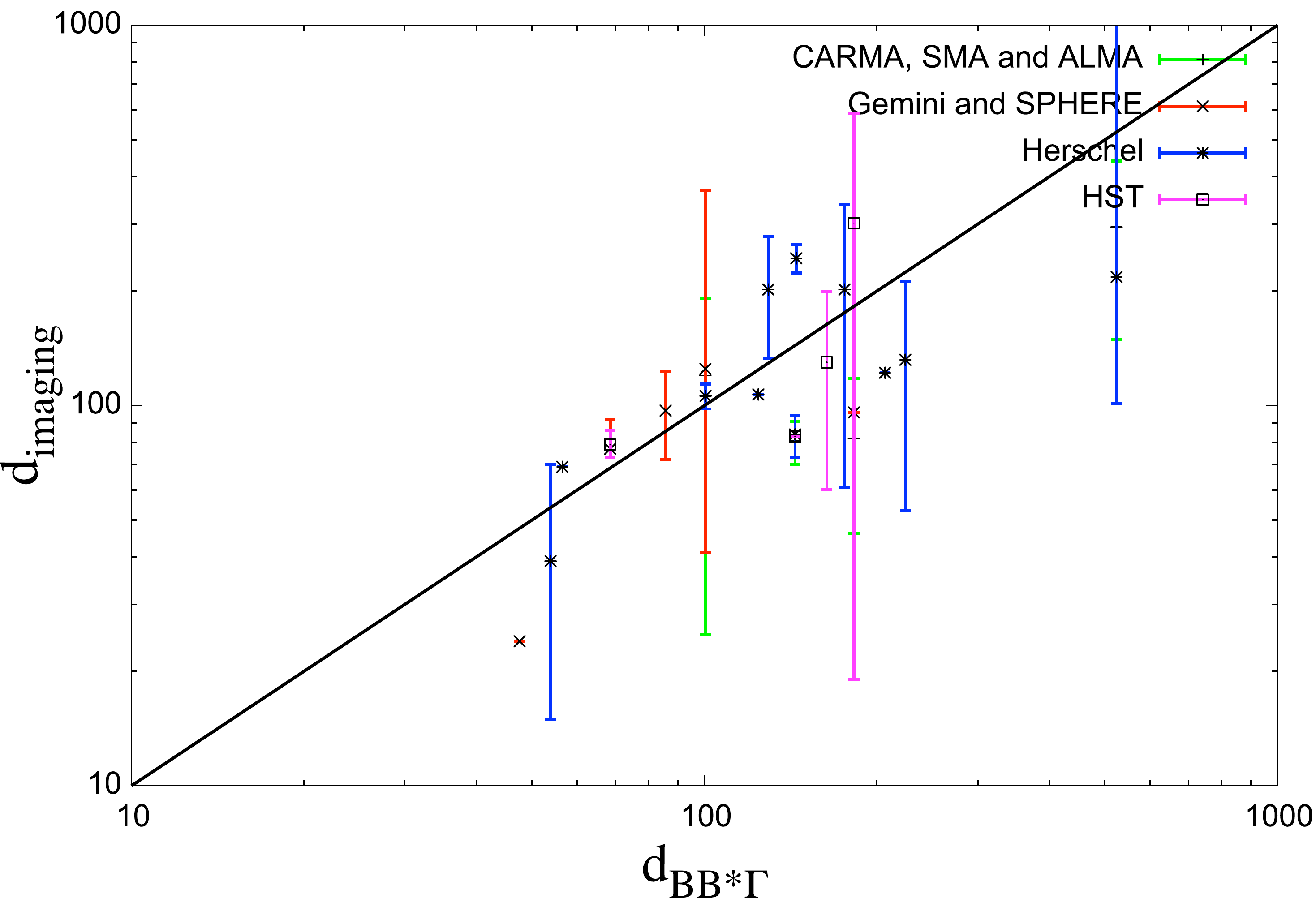}\\
\end{tabular}
\caption{Position of the outer belts as obtained from SED analysis with hypothesis of black-body (above) and with correction factor $\Gamma$ (beneath) versus positions obtained from resolved images. The black line represents the bisector, i.e. when the two positions coincide. The vertical error bars represent instead the extension of the disk from its inner edge to its outer one.}
\label{d2d2}
\end{figure}

From images of resolved disks we know that outer components are placed further away compared to the predicted black-body positions \citep{Pawellek1}. In \cite{Pawellek2} they analyzed a sample of 32 systems resolved by \textit{Herschel}/PACS and found a relation between the real radius of the disks and the black-body radius.\\
Defining the ratio between the two radii as $\Gamma$, they found that it depends on a certain power law of the luminosity of the star, 
\begin{equation}
\Gamma = A (L/L_{\odot})^B.
\label{pavell}
\end{equation}  
They explored five different compositions of dust grain: $50\%$ astrosilicates and $50\%$ vacuum,  $50\%$ astrosilicates and  $50\%$ ice,  $100\%$ astrosilcates,  $50\%$ astrosilicates and  $50\%$ carbon and  $100\%$ carbon. With the exception of  $100\%$ astrosilicates particles for which $A=8.26$ and $B=-0.55$,  they obtained similar values of A and B for each combination. We thus exclude pure astrosilicates and take mean values of the two parameters between the remaining compositions, $A=6$ and $B=-0.4$.\\
The use of the $\Gamma$ factor would be misleading for disks with high asymmetries or particular features such as extended halos that were, indeed, not considered by \cite{Pawellek2} in their analysis. Moreover, the $\Gamma$ correction is suitable only for systems with luminosity $L_*\ge L_{\odot}$.\\
\cite{Morales1} reach similar conclusions, showing for a sample of resolved systems with \textit{Herschel} that the position of the disk is better reproduced modeling the SED with dust grains composed of astrosilicates made of a mixture of dirty ice and pure water.\\ 
We show in Figure \ref{d2d2} how much the results of SEDs analysis with black-body hypothesis (upper panel) and with correction $\Gamma$ (lower panel) differ from data available in $17$ (out of $19$, we have to discard HD61005 and HD202917 because their luminosities are $< L_{\odot}$) resolved systems for the outer belt. The error bars represent the extension of the disks from the inner edge $R_{in}$ at the lower extremity to the outer edge $R_{out}$ at the upper one. Some systems have disks with small error bar because they are best modeled by thin rings centered on the mid-radii of the belts\\
In the upper panel, we can see a consistent difference and a systematic upward shift between SED modeling and direct imaging data, with the black-body fitting placing indeed the belt nearer to the star. This should not be a surprise since, as mentioned before, we expect that the disk is placed farther out when the particles do not behave like perfect absorbers/emitters. For these systems the situation improves applying the $\Gamma$ correction, as shown in the lower panel.\\
We note that also some error in the SED fitting could exist. Indeed, dust grains placed so far from the star have low temperatures (thus longer wavelengths) that are less constrained and more difficult to determine and only a few photometric points are available. Moreover, discordances between SED and direct imaging analysis could be caused by other factors. For example, the resolved systems are analyzed at certain wavelengths depending on the instrumentation used and the disk will appear quite different for each value of $\lambda$ whereas SED fitting identifies only the dust component. However, the fact that the \cite{Pawellek2} correction, developed for a sample of targets observed with \textit{Herschel}, holds reasonably well for a sample of objects resolved with different instruments/wavelengths indicate that such effects are not dominant in our sample. \\

\section{Spatially Resolved systems}

We list in Table \ref{tableA} the systems among the ones of our sample that were previously resolved in their farthest component by means of direct imaging. For each of such objects we show the instrument and the wavelengths at which the disk was resolved, the position $a$ of the mean radius of the disk and its inner and outer edges $R_{in}$ and $R_{out}$, the inclination $i$ (measured from face on in which case $i=0^{\circ}$) and the position angle $P.A.$. We note that the distances of the stars used in different papers may vary relative to each other and to distances used in this paper. For this reason we normalized the dimensions of the disks with distances listed in Table \ref{tabu1}. The only caveat is that this procedure should be entirely correct only when disk parameters are obtained directly from images and not from further modeling (such as SED modeling). 
\begin{table*} 
\tiny
\caption{Spatially Resolved systems}
\label{tableA}
\centering
\begin{threeparttable}
\begin{tabular}{c c c c c c c c c}
\hline\hline
Name & Instrument & $\lambda$ ($\mu m$) &a (AU)& $R_{in}$ (AU) & $R_{out}$ (AU) & $i (^{\circ})$ & $P.A. (^{\circ})$ & Reference\\
\hline
HD15115 & LBT/PISCES & 3.8 & (...) & 48 & 96E/120W &87  &275& \cite{Rodigas} \\
HD15115& HST/STIS &Visible &(...) &$\le19$ & 336E/586W& nearly edge-on& 30& \cite{Schneider1}\\
HD15115& Gemini/NICI & Near IR&96&(...)&(...)&86.2&98.5&\cite{Mazoyer}\\
HD15115& Subaru/IRCS& 1.63&(...)&92&337E/589W &86.3&278.63&\cite{Sai}\\
HD15115&SMA&1300&(...)&46&118&87&278.5&\cite{MacGregor}\\
HD30447 & HST/NICMOS  & Visible & (...) & 60 & 200 &nearly edge-on & 35& \cite{Soummer}\\
HD61005 & HST/STIS &Visible& 147 & 38 & 256 & 85.1 &70.3 &  \cite{Schneider1} \\
HD61005&HST/ACS&0.6&(...)&$\le33$&110&80.3&71.7&\cite{Maness}\\
HD61005&HST/NICMOS&1.1&(...)&$\le11$&223&(...)&160&\cite{Hines}\\
HD61005 & SMA& 1300& 73&(...)  &(...) & 70.3 & 84.3& \cite{Steele}\\
HD61005 & Herschel&70, 100, 160& 96& 90&101 &67  & 65& \cite{Morales1}\\
HD61005&VLT/NACO&1.65&65&65&149&84.3&70.3&\cite{Buenzli}\\
HD61005&VLT/SPHERE, ALMA& Near IR, 1300 &71&(...)&(...)&84.5&70.7&\cite{Olofsson}\\
HD61005&Gemini/GPI, Keck/NIRC2& Near IR& 48& 42&121&80&70.7&\cite{Esposito}\\
HD71155 &Herschel/PACS & 70, 100& 69 & (...)&(...) & 56.7&167.7 & \cite{Booth}\\
HD95086 & Herschel/PACS &70, 100, 160&202 & 61 & 338 &25.9 & 98.3& \cite{Moor} \\
$\beta$ Leo & Herschel/PACS & 100,160 & 39 & 15 & 69 & 35  & 125 & \cite{Churcher}\\
HD106906  &VLT/SPHERE&Near IR & 72 & 72 & 123 & 85 & 104& \cite{Lagrange} \\
HD106906& Gemini/GPI& Visible&(...)&56&>559&85&284&\cite{Kalas}\\
HR4796 & HST/STIS &Visible& 79 & 73 & 86 &76 & 27& \cite{Schneider} \\
HR4796&Gemini/NICI& Near IR&(...)&71&87&26&26.47&\cite{Wahhaj2}\\
HR4796&VLT/SPHERE&Near IR&77&73&92&76.4&27&\cite{Milli2}\\
HR4796&VLT/NACO&Near IR&78&71&84&75&26.7&\cite{Lagrange2}\\
HR4796&Keck/MIRLIN&12.5, 24.5&76&72&87&(...)&(...)&\cite{Wahhaj3}\\
HR4796&Keck II/OSCIR&10.8, 18.2&76&49&85&77&26.8&\cite{Telesco}\\
HR4796&HST/NICMOS&Visible&76&65&85&73.1&26.8&\cite{Schneider4}\\
HR4796&Magellan/MagAO& Near IR& 79&74&85&76.47&26.56&\cite{Rodigas2}\\
$\rho$ Vir &Herschel/PACS &70, 100, 160& 106 & 98&114 & 70 & 94& \cite{Morales1} \\
HD131835  &Gemini/TReCS &11.7, 18.3& 125 & 41 &368 &75  & 61& \cite{Hung}\\
HD131835 & Gemini/GPI& Visible&(...) &89&249&75.1&61.4&\cite{Hung2}\\
HD131835& ALMA & 1240& 120&25&191&73&58&\cite{Lieman}\\
HD131835 \tnote{a}& VLT/SPHERE& Near IR&114 & 89 &166&72.6& -120& \cite{Feldt}\\
HD141378 & Herschel/PACS &100, 160&202 &133 &279 &62 &113 & \cite{Morales1}\\
$\pi$ Ara & Herschel/PACS &100, 160& 122 &(...) & (...)&40 &3 & \cite{Morales1} \\
$\eta$ Tel & Gemini/TReCS & 11.7, 18.3 & 24 &(...) &(...) & 83 &8 & \cite{Smith}\\
HD181327 & HST/STIS &Visible& 83 &(...) &(...) & 30.1 &102 & \cite{Schneider1} \\
HD181327&HST/NICMOS& Visible&83&66&100&31.7&107&\cite{Schneider3}\\
HD181327& Herschel/PACS &  70, 100, 160& 84&73&94&31.7&107& \cite{Lebreton}\\
HD181327& ALMA& 1300&81&70&91&30&98.9&\cite{Marino}\\
HD188228 & Herschel/PACS &70, 100& 107 & (...)&(...) & 34.3 & 11.4& \cite{Booth} \\
$\rho$ Aql & Herschel/PACS &100, 160&244 & 223& 265&68 & 93& \cite{Morales1} \\
HD202917 & HST/STIS &Visible& 69 & 61 & 76 &  68.6 & 108& \cite{Schneider2}\\
HD202917&HST/NICMOS&Visible&(...)&(...)&118&70&300&\cite{Soummer}\\
HD206893& Herschel/PACS& 70&(...)&53&212&40&60&\cite{Milli}\\
HR8799 & Herschel/PACS &70, 100, 160, 250&  218 & 101 & 2000 & 26 & (...)& \cite{Matthews} \\
HR8799& ALMA& 1340&(...)&149&440&40&51&\cite{Booth2}\\
HR8799&SMA&880&172&152&304&20&(...)&\cite{Hughes}\\
HR8799&Spitzer& 40, 70 &(...)&92&308&$\le25$&(...)&\cite{Su3}\\
HR8799&CSO& 350&(...)&102&304&(...)&(...)&\cite{Patience}\\
\hline
\end{tabular}
\begin{tablenotes}
\item[a] HD131835 shows a multiple rings structure between the belts at $\sim 115$ AU and $\sim 6$ AU and this could be an indication of the presence of (still forming) planets. Such internal structures, however, are not well constrained and they may be short-living. Therefore, we will treat this system as all the others in the sample noticing that further and more precise analysis could be done but it is beyond the aims of this paper.
\end{tablenotes}
\end{threeparttable}
\end{table*}

\end{appendix}

\end{document}